\documentclass[%
 reprint,
 pre,
 amsmath,amssymb,
 aps,
]{revtex4-2}

\usepackage{graphicx}
\usepackage{dcolumn}
\usepackage{bm}

\usepackage{xcolor}

\usepackage{hyperref}

\begin{document}

\preprint{APS/123-QED}

\title{Thermodynamic Bounds from Otto--Villani Functional Inequalities}

\author{Andrea Auconi}
\email{andrea.auconi@gmail.com}

  \affiliation{%
 Ca’ Foscari University of Venice, DSMN - via Torino 155, 30172 Mestre (Venice), Italy
}%

\date{\today}

\begin{abstract}
The dissipation in the relaxation of an ensemble of conservative stochastic systems toward the {equilibrium} steady state is quantified by the free energy difference. Functional inequalities within the framework of [F. Otto and C. Villani, J. Funct. Anal. \textbf{173}, 361 (2000)] are here revisited which connect the free energy dynamics and optimal transport, offering a geometric perspective on the instantaneous speed of relaxation in the presence of potential barriers. These are illustrated with numerical relaxation experiments on Landau-Ginzburg potentials.
{The same inequalities hold for relaxation toward nonequilibrium steady states once the potential is replaced by the pseudo-potential defined by the stationary density.}
\end{abstract}

\maketitle

\section{Introduction}

In nonequilibrium thermodynamics \cite{kubo2012statistical,gaspard2022statistical,seifert2012stochastic}, a system's response to sudden changes in external conditions fundamentally manifests as transient relaxation dynamics. 
These relaxation processes are found across various fields, ranging from sensory adaptation in biophysics \cite{lan2012energy, ito2015maxwell,edelson2025context} and bit erasure in information processing \cite{berut2012experimental,ProesmansFinite2020,lee2022speed,oikawa2025experimentally}, to Mpemba effects in thermal phenomena \cite{lu2017nonequilibrium, teza2026speedups,lapolla2020faster,meibohm2021relaxation,ibanez2024heating}.

For continuous systems, relaxation processes are typically described at the ensemble level in terms of the Fokker-Planck equation \cite{risken1989fokker,jordan1998variational,ito2024geometric}, and the time derivative of the relative entropy from the steady state is a measure of instantaneous relaxation speed.
This has the mathematical form of a relative Fisher information \cite{villani2009optimal}, and the thermodynamic interpretation of an excess entropy production rate \cite{hatano2001steady,van2010three,esposito2011second,dechant2022geometric,kolchinsky2026generalized}.

Thermodynamic speed limits formalize the trade-off between the entropy production in a finite-time process and its duration \cite{PhysRevLett.121.070601,shiraishi2019information}.
In this framework, the Benamou-Brenier formulation of optimal transport \cite{benamou2000computational} has been considered as a variational principle to derive speed limits in terms of the Wasserstein distance \cite{aurell2012refined,dechant2019thermodynamic,nakazato2021geometrical,van2023thermodynamic,sabbagh2024wasserstein,shiraishi2024wasserstein,zhong2024beyond}.
These speed limits are inherently defined for finite-time processes, as indeed when applied to bound the instantaneous relaxation speed they either {reduce to the definition of the entropy production rate}, or simply vanish.


The Otto--Villani HWI inequality \cite{otto2000generalization,lott2009ricci,gentil2020entropic,karatzas2022trajectorial} offers a lower bound on the instantaneous relaxation speed based on the Wasserstein distance and the minimum convexity of the potential.
Although the HWI bound is tighter than the Logarithmic Sobolev inequality (LSI), this worst-case scenario on the potential landscape makes it typically loose in non-convex potentials. In contrast, this paper highlights an intermediate bound in the HWI derivation that provides the fundamental connection between physical {density dynamics} and optimal transport, and offers a geometric perspective on {thermal relaxation}. 
After reviewing the functional inequalities with optimal transport, the predictive power of the various bounds is illustrated analytically on a 1D Gaussian expansion and numerically on a double-well potential.

{The exposition and examples considered deal with conservative systems relaxing toward an equilibrium steady state. The considered inequalities hold in nonconservative systems relaxing toward a nonequilibrium steady state once the potential is replaced by the pseudo-potential defined by the stationary density, in which case they bound the excess entropy production rate.}

\section{Physical flow and optimal transport}

\paragraph*{Fokker-Planck equation.}

Consider an overdamped Brownian particle at position $\boldsymbol{x}$ in $n$-dimensional space, being in contact with a heat bath at temperature $T$ and subject to a potential $\psi\equiv \psi(\boldsymbol{x})$.
Its stochastic dynamics is described by the equilibrium Langevin equation \cite{seifert2012stochastic},
\begin{equation}\label{Langevin}
    d\boldsymbol{x} = - \boldsymbol{\nabla} \psi \,dt + \sqrt{2T} d\boldsymbol{W},
\end{equation}
where $d\boldsymbol{W}$ are standard vector Brownian motion increments \cite{karatzas2014brownian}, and the mobility and Boltzmann constant are set to unity and not considered.
The corresponding probability density $p\equiv p(\boldsymbol{x},t)$ evolves according to the Fokker-Planck equation \cite{risken1989fokker,jordan1998variational,ito2024geometric},
\begin{equation}\label{FP}
    \partial_t p = \boldsymbol{\nabla}\cdot \left[ p \boldsymbol{\nabla}(\psi + T\ln p ) \right],
\end{equation}
where $-\boldsymbol{\nabla}(\psi + T\ln p )$ is the local mean velocity.
Assume the potential $\psi$ to be such that a unique invariant density exists defined by $\partial_t p|_{p^*}= 0$. The latter is then given by the Gibbs-Boltzmann distribution
\begin{equation}
    p^* = \frac{1}{Z} \exp\left(-\frac{\psi}{T}\right) ,
\end{equation}
where $Z\equiv \int d\boldsymbol{x} \, \exp\left(-\psi/T \right)$ ensures normalization.
Quantities evaluated at the steady state are marked with an asterisk ($*$).
In terms of the steady-state density one can rewrite $\boldsymbol{\nabla}(\psi + T\ln p ) = T \boldsymbol{\nabla} \ln (p/p^*) $.

\paragraph*{Free energy dynamics.}
The Helmholtz free energy \cite{seifert2012stochastic} is defined as
\begin{equation}
    \mathcal{F} \equiv U-TS ,
\end{equation}
where $U[p]\equiv \int d\boldsymbol{x} \, p \psi$ is the potential energy, and $S[p] \equiv -\int d\boldsymbol{x} \, p \ln p$ is Shannon entropy.
The equilibrium free energy reads $\mathcal{F}^*= -T \ln Z$.

The relative entropy \cite{amari2016information} of a distribution $p\equiv p(\boldsymbol{x})$ from a reference distribution $q\equiv q(\boldsymbol{x})$ is defined as $D[p\| q]\equiv \int d\boldsymbol{x}\,p \ln\left( p/q \right)\geq 0$, and it is {nonnegative} by Jensen's inequality.
The relative entropy from the invariant density is related to the free energy \cite{hatano2001steady,van2010three,esposito2011second,dechant2022geometric,kolchinsky2026generalized} as
\begin{equation}\label{def D}
    D\equiv D[p\| p^*] 
    = \frac{1}{T}\left( \mathcal{F}-\mathcal{F}^* \right) \geq 0 .
\end{equation}
Since $\mathcal{F}^*$ is constant, the dynamics of the relative entropy is proportional to that of the free energy, $T d_t D = d_t \mathcal{F}$, 
which yields
\begin{equation}\label{d_t D}
    d_t D = - T \int d\boldsymbol{x}\, p \, \left\| \boldsymbol{\nabla}  \ln \left(\frac{p}{p^*}\right) \right\| ^2 \leq 0 ,
\end{equation}
whose negativity guarantees the Glansdorff-Prigogine stability of the steady state \cite{glansdorff1974thermodynamic,ito2022information, maes2015revisiting}.
This relation of Eq. \eqref{d_t D} between a relative Fisher information and the time derivative of a relative entropy is a type of De Bruijn's identity \cite{dembo2002information}.
The quantity $-d_t D$ is also known as the Hatano-Sasa excess entropy production rate 
\cite{hatano2001steady,van2010three,esposito2011second,dechant2022geometric,kolchinsky2026generalized}.
Note that the restriction to a purely conservative 
dynamics Eq. \eqref{Langevin} does not explicitly impact $d_t D$, as indeed nonconservative forces only enter at the second time derivative $d^2_t D$, see Refs. \cite{auconi2025nonequilibrium,auconi2026information}.

\paragraph*{Optimal transport.}
The Wasserstein distance $\mathcal{W}$ quantifies the minimum quadratic cost of transforming one probability distribution into another \cite{villani2009optimal},
\begin{equation}\label{def W}
  \mathcal{W}^2(p, q) \equiv \inf_{\pi \in \Pi(p, q)} \int d\boldsymbol{x} d\boldsymbol{y} \, \pi(\boldsymbol{x}, \boldsymbol{y}) \| \boldsymbol{x} - \boldsymbol{y}\| ^2,
\end{equation}
where $\Pi(p, q)$ is the set of all joint probability distributions $\pi(\boldsymbol{x}, \boldsymbol{y})$ whose marginals are the given densities $\int d\boldsymbol{y} \, \pi =p(\boldsymbol{x})$ and $\int d\boldsymbol{x} \, \pi = q(\boldsymbol{y})$.

This optimal transport problem admits an equivalent formulation, due to Benamou and Brenier \cite{benamou2000computational}, as the minimization of the time-integrated kinetic energy of a fluid having the given densities as temporal boundary conditions.
Such optimal fluid dynamics, reviewed in the Supplementary Materials (SM) which includes Refs. \cite{villani2021topics,dechant2018entropic,ito2020stochastic,ekeland1999convex, bao2025universal}, is written in terms of the time-dependent velocity field {$\boldsymbol{\nabla}\lambda_\tau\equiv \boldsymbol{\nabla}\lambda_\tau(\boldsymbol{x})$} as
\begin{equation}\label{optimal transport}
\begin{cases}
  {\partial_\tau} p_\tau = - \boldsymbol{\nabla} \cdot (p_\tau \boldsymbol{\nabla}\lambda_\tau), \\
  {\partial_\tau} \lambda_\tau = -\frac{1}{2} \| \boldsymbol{\nabla} \lambda_\tau\| ^2 ,
\end{cases}
\end{equation} 
where {$\tau$ is a fictitious time variable which parametrizes the optimal transport path $p_\tau$ in $\tau \in [0, 1]$.} 
As seen from the total derivative $d_\tau\boldsymbol{\nabla}\lambda_\tau(\boldsymbol{x}_\tau)=\boldsymbol{0}$ {along the trajectory $\boldsymbol{x}_\tau$ defined by $d_\tau\boldsymbol{x}_\tau=\boldsymbol{\nabla}\lambda_\tau(\boldsymbol{x}_\tau)$,} Eq.~\eqref{optimal transport} describes fluid particles travelling on straight lines at constant velocity. Accordingly, the  Wasserstein distance can be evaluated from the field $\lambda_\tau$ at any transport time point as $\mathcal{W}^2(p_0, p_1) =\int d\boldsymbol{x}\, p_\tau \,\|  \boldsymbol{\nabla}\lambda_\tau \| ^2 $, and the optimal map $\boldsymbol{T}$ that transforms the initial distribution $p$ to the final target $q$ is written $\boldsymbol{T}(\boldsymbol{x}) = \boldsymbol{x} + \boldsymbol{\nabla} \lambda_0(\boldsymbol{x})$.

\begin{figure}[ht]
    \centering
    \includegraphics[width=\columnwidth]{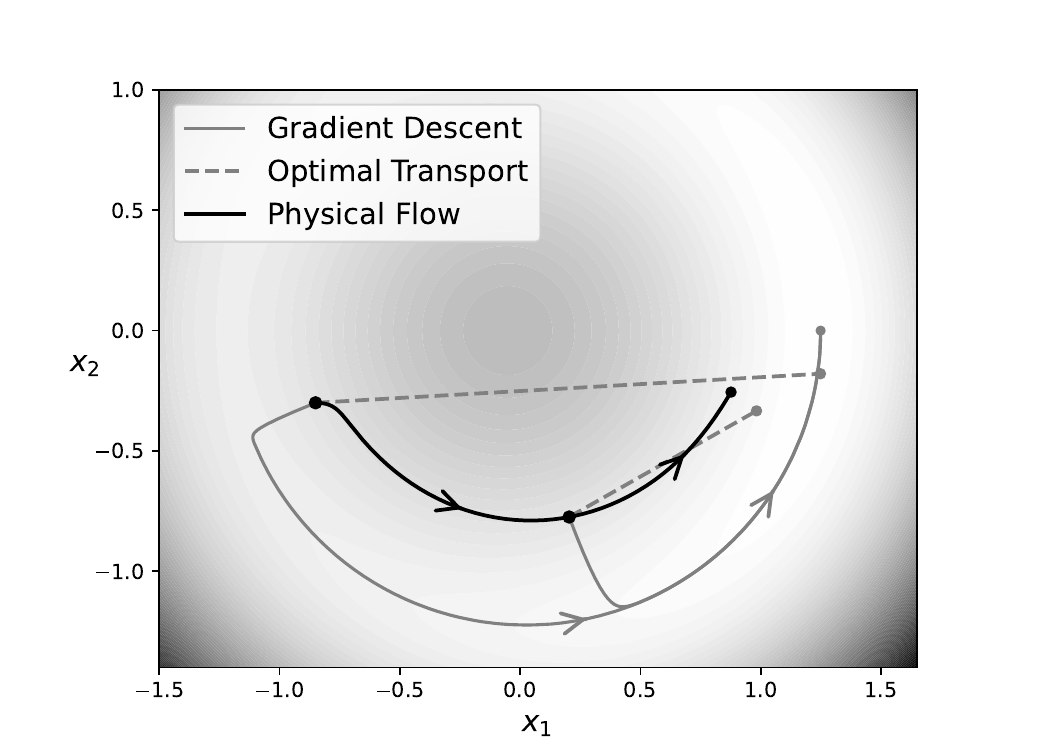}
    \includegraphics[width=\columnwidth]{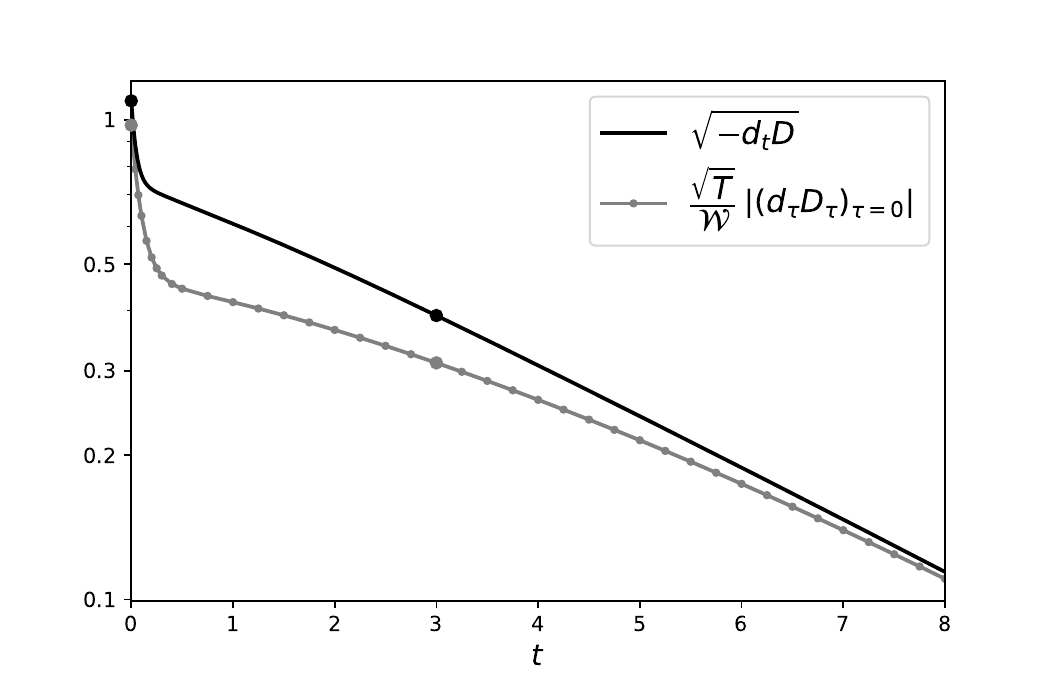}
    \caption{The gradient descent velocity is $-\boldsymbol{\nabla}\psi$, based only on the potential. 
    The physical flow is defined as the local mean velocity $- \boldsymbol{\nabla} (\psi + T\ln p)$, which also accounts for the temperature-driven diffusion.
    The corresponding optimal transport ensemble $p_\tau$ is made of fluid particles of constant velocity $\boldsymbol{\nabla}\lambda_{\tau}(\boldsymbol{x}_{\tau}) $.
    Consider a tilted Mexican hat potential, $\psi(\boldsymbol{x}) = \boldsymbol{z}\cdot \boldsymbol{x} - \alpha \| \boldsymbol{x} \|^2 + \beta \| \boldsymbol{x} \|^4$, with $\alpha = 3 $, $\beta = 1$, $T=0.3$, and $\boldsymbol{z} = 0.3 \boldsymbol{\hat{x_1}}$ along the $x_1$ axis, and an ensemble brought out of equilibrium at $t=0$ by switching the bias as $\boldsymbol{z}\rightarrow -\boldsymbol{z}$. Numerical integration of the physical flow shows how trajectories typically bend in the presence of a potential barrier along the geodesic. {Gradient descent and optimal transport are shown from two points along the same physical trajectory, as at each time $t$ the current density $p(t)$ defines its own optimal transport path toward $p^*$.
    The bottom panel compares the instantaneous relaxation speed $\sqrt{-d_t D}$ (black line) to the lower bound of Eq.~\eqref{general bound} (gray line). 
    The bound is least tight shortly after the switch and local relaxation, when the physical flow is redirected along the valley around the central peak. It approaches saturation at later times, when a single slow relaxation mode dominates and the density is close to the steady state $p\simeq p^*$.
    Neither the HWI [Eq.~\eqref{HWI}] nor the LSI [Eq.~\eqref{LogSobolev}] provides a positive bound here.}}
    \label{illustration}
\end{figure}

Finding the particular initial field configuration $\lambda_0(\boldsymbol{x})$ that solves this boundary value problem is approached with numerical techniques like the entropically regularized optimal transport via the Sinkhorn-Knopp algorithm \cite{peyre2019computational}. 
See Fig. (\ref{illustration}) for an illustration of the optimal transport in comparison to gradient descent and to the physical flow defined by the local mean velocity.


\section{Thermodynamic bounds in the Otto--Villani framework}

\paragraph*{The general bound.}

Consider the optimal transport from the current density $p$ at physical time $t$ to the steady-state density $p^*$.
{Let us introduce the notation $f_{\tau} \equiv f(\boldsymbol{x}, t, \tau)$ for any spatial field which depends on the transport time $\tau$, and use the convention $f \equiv f_0$ for the initial transport time $\tau=0$.
Similarly for the relative entropy functional we define $D_{\tau}\equiv D[p_\tau \|p^*]$ and $D \equiv D[p \|p^*]$.
Note that for any particular configuration of the physical density $p$ at any time $t$, there exists an optimal transport dynamics $p_\tau$ described by $\lambda_\tau\equiv\lambda(\boldsymbol{x}, t, \tau)$ through Eq. \eqref{optimal transport} with $\tau\in [0,1]$ having the steady-state density $p_1=p^*$ as target. We denote the corresponding Wasserstein distance as $\mathcal{W}\equiv \mathcal{W}(p, p^*)$. The notation adopted in this section reflects that we consider $t$ as a fixed parameter here, and study the corresponding optimal transport as a function of $\tau$.}

Under the optimal transport dynamics, described by Eq. \eqref{optimal transport}, the relative entropy evolves as
\begin{equation}\label{d_t D optimal transport}
    {d_\tau D_\tau = \int d\boldsymbol{x} \, p_\tau\, \boldsymbol{\nabla} \lambda_\tau \cdot \boldsymbol{\nabla} \ln \left(\frac{p_\tau}{p^*}  \right).}
\end{equation}
By applying the Cauchy-Schwarz inequality {at $\tau=0$ and recognizing the definition Eq. \eqref{d_t D}} one obtains the thermodynamic bound
\begin{equation}\label{general bound}
{\sqrt{-d_t D} \geq \frac{\sqrt{T}}{\mathcal{W}} \left|  (d_\tau D_\tau)_{\tau=0} \right|  \geq 0,}
\end{equation}
which is the fundamental connection between optimal transport and physical relaxation as derived by Otto and Villani \cite{otto2000generalization,Note1}.
Its nonnegativity establishes it as a refinement of the second law of thermodynamics.
The equality case is obtained when physical local mean velocity $-T\boldsymbol{\nabla}\ln (p/p^*)$ and optimal transport velocity $\boldsymbol{\nabla}\lambda$ are globally proportional.
{Recall that} the notation {$(d_\tau D_\tau)_{\tau=0}$} here means the time derivative along the optimal transport path defined by $p$ and $p^*$, evaluated at the initial transport time {$\tau = 0$} where {the physical and optimal transport densities coincide $p_0=p$}. This implicitly means to consider the initial fluid velocity field {$\boldsymbol{\nabla} \lambda = \boldsymbol{\nabla} \lambda(\boldsymbol{x}, t, 0)$} solution of the optimal transport problem {at physical time $t$}.

This inequality provides a geometric perspective on the slowing down of relaxation in the presence of potential barriers, which is at the basis of the Markovian Mpemba effect \cite{lu2017nonequilibrium, teza2026speedups,lapolla2020faster,meibohm2021relaxation,ibanez2024heating}.
Consider a landscape where two or more wells are separated by large potential barriers. If a density is released within one of these wells, it first undergoes a rapid relaxation of timescale $t_\text{fast}$ until it locally reproduces the invariant shape, $p(t_\text{fast}) \approx \mathbb{I}_{\boldsymbol{x}\in \Omega}\, p^* /\int_{\Omega} d\boldsymbol{x} \, p^*$, where $\Omega$ denotes the corresponding region. Then for $t>t_\text{fast}$ a kinetic bottleneck is predicted, since both $d_t D$ and {$(d_\tau D_\tau)_{\tau=0}$} integrate the factor $p \boldsymbol{\nabla} \ln (p/p^*) \approx \boldsymbol{0}$, which is small in the wells as $\boldsymbol{\nabla} \ln (p/p^*)|_{\boldsymbol{x}\in \Omega} \approx \boldsymbol{0}$ and small on the peaks as $p|_{\boldsymbol{x}\notin \Omega} \approx 0$ for large barriers. 
Crucially, in $n\geq 2$ dimensions the physical dynamics may exploit alternative pathways around the barriers rather than following the direct optimal transport path, see Fig. (\ref{illustration}), rendering Eq. \eqref{general bound} a strict inequality.
{Note that, even in one dimension, where no alternative pathways exist,
Eq.~\eqref{general bound} is typically strict as the physical flow and the
geodesic in general point in different directions in the space of densities.}

{In other words, the right-hand side of Eq.~\eqref{general bound} is the relaxation speed that the ensemble would have if its local mean velocity were replaced by its projection onto the geodesic. This constrained speed is set by the slope of the free energy along the geodesic, $T(d_\tau D_\tau)_{\tau=0}$, which is reduced by the energetic term $\int d\boldsymbol{x}\,p\,\boldsymbol{\nabla}\lambda\cdot\boldsymbol{\nabla}\psi$ when the direct route has to climb a potential barrier. The physical flow, instead, deviates from the geodesic, for example around the barriers, and thereby relaxes faster than along the unfavourable direct route.}

{Let us note that, being a lower bound, Eq.~\eqref{general bound} cannot by itself imply a slowing down of relaxation. In fact, its gap quantifies the speed-up of the physical flow with respect to the direct transport route, while a small lower bound implies a slow relaxation only close to saturation.}

\paragraph*{HWI inequality.}

For any smooth function $f(s)$, the Taylor identity $f(1)=f(0)+ f'(0) +\int_0^1 ds\, (1-s) f''(s)$ holds, which applied to {$D_\tau$ with respect to the optimal transport time $\tau$} 
reads
\begin{equation}\label{integral relation}
{  -(d_\tau D_\tau )_{\tau=0}  =  D + \int_0^1 {d\tau \, (1-\tau) d^2_\tau } D_\tau,}
\end{equation}
{where we used $D_1 = D[p^* \| p^* ]=0$}, and the second derivative is computed as
\begin{equation}\label{2nd derivative optimal transport}
      {  d^2_{\tau} D_\tau 
   = \int d\boldsymbol{x} \, p_\tau \left( \boldsymbol{\nabla}\lambda_\tau \cdot \boldsymbol{\nabla}\boldsymbol{\nabla}\psi \cdot \boldsymbol{\nabla}\lambda_\tau /T + \|\boldsymbol{\nabla}\boldsymbol{\nabla}\lambda_\tau\|^2 \right) ,}
\end{equation}
where $\|\boldsymbol{\nabla}\boldsymbol{\nabla}\lambda_\tau\|^2$ denotes the Frobenius norm.


The minimum convexity of the potential $\psi$ is defined via the Loewner order as
\begin{equation}\label{kappa}
{\kappa \equiv \inf_{\boldsymbol{x}} \min_{\boldsymbol{u}} \frac{\boldsymbol{u} \cdot  \boldsymbol{\nabla}\boldsymbol{\nabla} \psi \cdot \boldsymbol{u}}{\| \boldsymbol{u}\|^2}} ,
\end{equation}
which means that the eigenvalues of its Hessian matrix $\boldsymbol{\nabla}\boldsymbol{\nabla} \psi$ are bounded below by $\kappa$ across the domain.
From this global majorization [Eq. \eqref{kappa}] and the nonnegativity $\|\boldsymbol{\nabla}\boldsymbol{\nabla}\lambda_\tau\|^2\geq 0$, the second derivative 
[Eq. \eqref{2nd derivative optimal transport}] is bounded by
${d^2_{\tau} D_\tau } \geq (\kappa/T) \mathcal{W}^2$ for all transport times $0\leq \tau \leq 1 $.
Substituting into Eq. \eqref{integral relation} and using the inequality Eq. \eqref{general bound} yields
\begin{equation}\label{HWI}
    \sqrt{-d_t D}  \geq \frac{\sqrt{T}}{\mathcal{W}}  D +  \frac{\mathcal{W}}{\sqrt{T}} \frac{\kappa}{2} ,
\end{equation}
which is the celebrated HWI inequality \cite{otto2000generalization}.
It depends on the optimal transport problem only through the Wasserstein distance $\mathcal{W}$, and it unifies other functional inequalities also based on the minimum convexity $\kappa$, like the LSI and Talagrand inequality. 
The equality case in the HWI is obtained if the potential has constant convexity $\boldsymbol{\nabla}\boldsymbol{\nabla} \psi = \kappa \boldsymbol{I}$ (meaning $p^*$ is a Gaussian), and $p$ is a pure translation of $p^*$.
Note that, while Eq. \eqref{HWI} holds for non-convex potentials with $\kappa < 0$, the right-hand side is positive only for $\kappa > - 2T D / \mathcal{W}^2 $, therefore the HWI inequality is not a general refinement of the second law of thermodynamics.

Let us discuss the three majorization steps taken to obtain the HWI inequality from the intermediate bound Eq. \eqref{general bound}.
First, the absolute value majorization {$\left| ( d_\tau D_\tau )_{\tau=0} \right| \geq - ( d_\tau D_\tau )_{\tau=0}$} restricts the {usefulness} to regimes where the relative entropy decreases along the geodesic {at $\tau =0$}. Then, the minimum convexity majorization $\boldsymbol{\nabla}\lambda_\tau \cdot \boldsymbol{\nabla}\boldsymbol{\nabla}\psi \cdot \boldsymbol{\nabla}\lambda_\tau\geq \kappa \| \boldsymbol{\nabla}\lambda_\tau \|^2$ effectively compares the state to the worst-case scenario of having to cross the sharpest peak in the landscape. Finally, the Frobenius norm majorization neglects the entropic impact of expansions and contractions, as $\boldsymbol{\nabla}\boldsymbol{\nabla}\lambda_\tau$ is the spatial gradient of the optimal transport velocity field $\boldsymbol{\nabla}\lambda_\tau$.
Note that dimensional refinements to this term have been obtained \cite{erbar2015equivalence} by the trace inequality $\| \boldsymbol{\nabla}\boldsymbol{\nabla}\lambda_\tau\| ^2\geq (\nabla^2 \lambda_\tau)^2 /n$, see also the SM.


\begin{figure}[ht]
    \centering
    \includegraphics[width=\columnwidth]{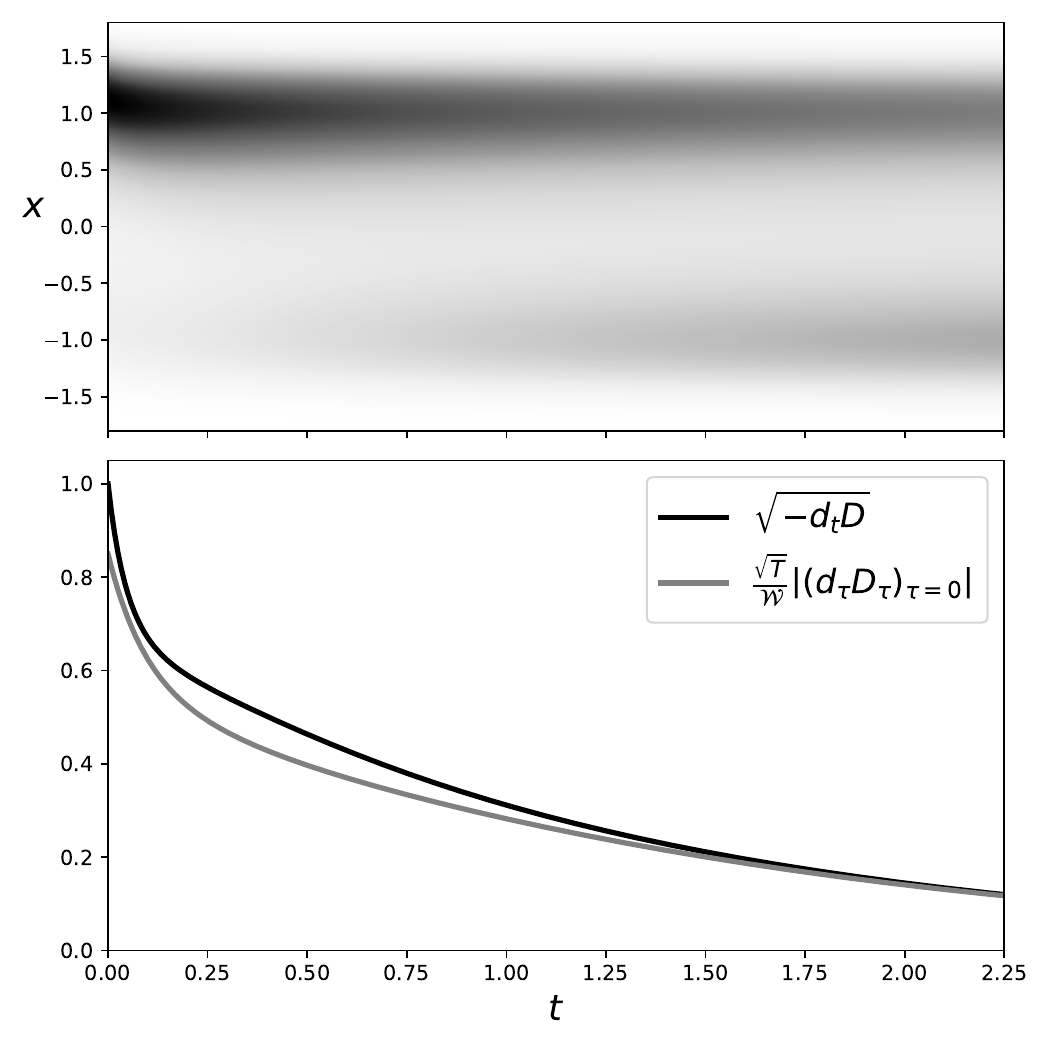}
    \caption{{Thermal decay of a stored bit (reverse of bit erasure)} in a Landau-Ginzburg double-well potential $\psi = -\alpha x^2 + \beta x^4$. The parameters are set to $T=1$, $\alpha=2$, $\beta=1$, with an initial confining bias $\gamma=-1$. (Top) Relaxation of the probability density $p(x,t)$ over the central energy barrier toward a symmetric bimodal steady state. (Bottom) The instantaneous relaxation speed $\sqrt{-d_t D}$ (black line) compared to the lower bound of Eq. \eqref{general bound} (gray line) based on optimal transport.
    Note the acceleration change around the local intra-well relaxation timescale $t_\text{fast}=1/(4\alpha)$, after which the dynamics is dominated by the slow inter-well relaxation.   
    The HWI Eq. \eqref{HWI} does not provide a positive bound here as in the considered time interval it holds uniformly $\kappa < - 2T D / \mathcal{W}^2 $.}
    \label{bit_erasure}
\end{figure}

\paragraph*{Logarithmic Sobolev inequality.}

One may be interested in a thermodynamic bound independent of the optimal transport problem. This is the case of the LSI, which could be derived either directly from the physical dynamics as reviewed in the SM, or as a consequence of the HWI inequality as done in \cite{otto2000generalization}.
Indeed, by taking the minimum of the right-hand side of Eq. \eqref{HWI} with respect to the Wasserstein distance $\mathcal{W}$ as if it were a free parameter, it yields the LSI
\begin{equation}\label{LogSobolev}
    - d_t D  \geq 2 \kappa D  .
\end{equation}
Clearly this inequality becomes trivial if $\kappa < 0$, meaning when the potential is not convex everywhere.
When $\kappa > 0$, the LSI guarantees the exponential convergence to the steady state as $D[p(t) \| p^*] \leq D[p(0) \| p^*] \exp(-2\kappa t)$.

\paragraph*{Nonequilibrium steady states.}

{While the derivations above consider conservative dynamics relaxing to an equilibrium Gibbs-Boltzmann distribution, the geometric framework naturally extends to systems driven by nonconservative forces. Let us first note that, in a nonequilibrium steady state, the presence of persistent probability currents does not explicitly impact the instantaneous relaxation speed measured by $d_t D$, see, e.g.,  Refs. \cite{auconi2025nonequilibrium,auconi2026information}. We also note that the optimal transport map depends solely on the spatial boundary densities $p$ and $p^*$, independently of the underlying physical fluxes. Consequently, the intermediate bound of Eq. \eqref{general bound}  applies equally to relaxation toward an equilibrium or a nonequilibrium steady state. The subsequent HWI and LSI inequalities [Eqs. \eqref{HWI}-\eqref{LogSobolev}] also remain formally valid for nonequilibrium systems, with the required modification that the minimum convexity should be evaluated on the pseudo-potential $\psi_{\text{eff}} \equiv -T \ln p^*$ defined by the steady-state density.}

\begin{figure}[t]
    \centering
    \includegraphics[width=\columnwidth]{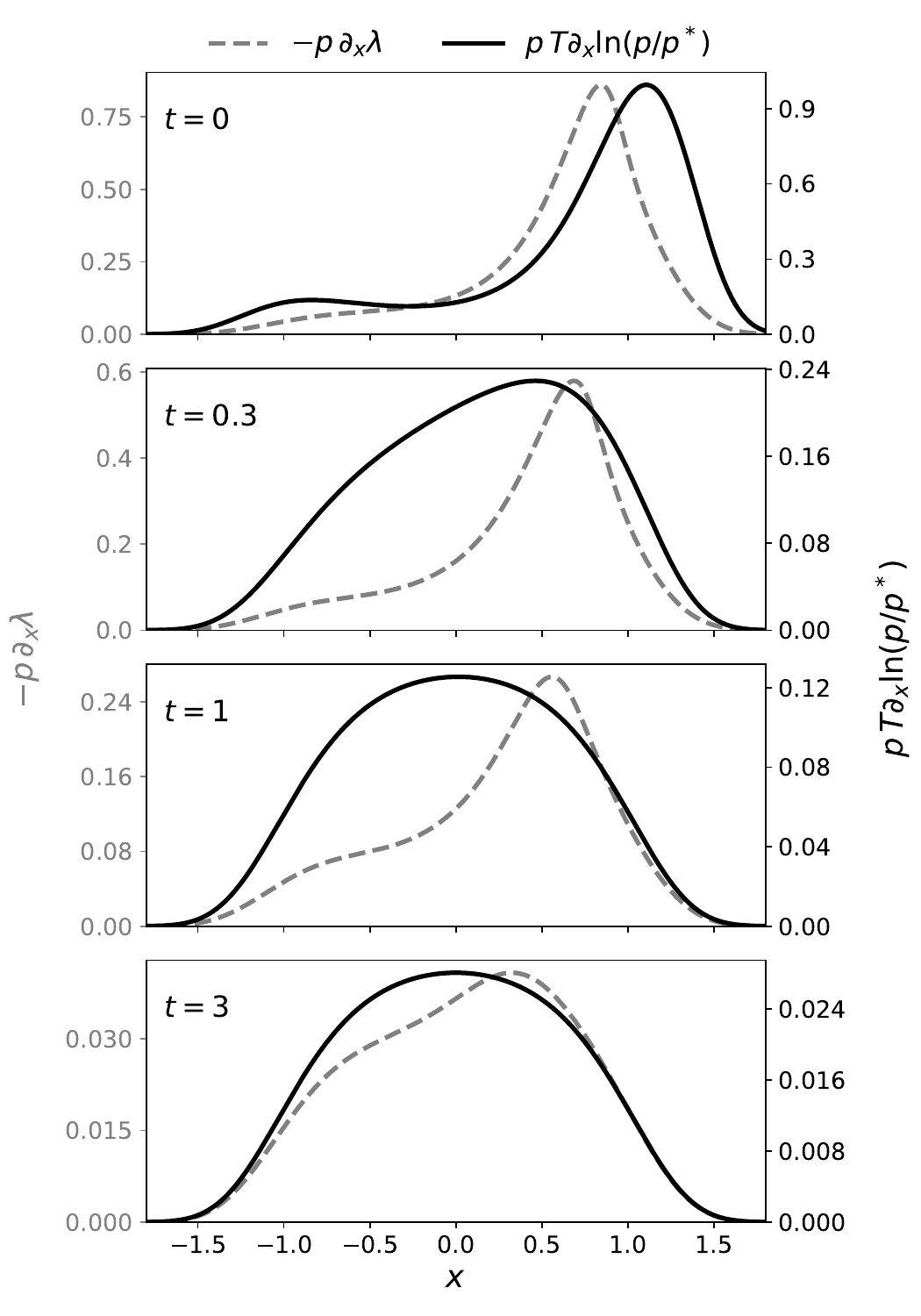}
    \caption{{Probability currents during the relaxation of Fig.~\ref{bit_erasure}. The optimal transport flux $p\,\partial_x\lambda$ (gray dashed line) and the physical current $-pT\partial_x\ln(p/p^*)$ (black line), both with inverted sign and on separate vertical scales, so that proportional currents overlap.} 
    }
    \label{bit_erasure_fields}
\end{figure}

\section{Numerical illustrations}

\paragraph*{Illustration in a double-well potential.} 
Consider the Landau-Ginzburg double-well potential in 1D, $\psi(x, t\geq 0) = -\alpha x^2 + \beta x^4$, with $\alpha >0$ and $\beta>0$.
The ensemble is initially in thermal equilibrium with respect to a tilted version of the potential, $\psi(x, t<0) = \psi(x, t\geq 0) + \gamma x$, where the bias $\gamma<0$ forces the probability mass into the right well. At time $t=0$, the bias is removed and the system relaxes toward the symmetric bimodal equilibrium distribution $p^* \sim \exp(-\psi(x,t\geq 0)/T)$. This transition from a localized state to a symmetric delocalized state across an energy barrier is analogous to the {reverse} process of bit erasure \cite{berut2012experimental,ProesmansFinite2020,lee2022speed,oikawa2025experimentally}.
Note that the one-dimensional scenario is uniquely tractable as the optimal transport map can be constructed through the monotonic matching of cumulative distributions \cite{rioul2017optimal}. 
If $F$ and $F^*$ are smooth and strictly increasing cumulative distributions, the optimal transport map from $F$ to $F^*$ is given by the composition $(F^*)^{-1} \circ F$. 
As demonstrated in Fig. \ref{bit_erasure}, the fundamental bound Eq. \eqref{general bound} becomes tight only in the long-time limit. 
At early times $t<t_\text{fast}=1/(4\alpha)$ the physical flow and optimal transport fields are not globally proportional due to the simultaneous presence of fast intra-well and slow inter-well relaxation. Once the intra-well dynamics reaches a metastable local equilibrium, the only surviving process is the slow leakage of mass across the central energy barrier. 
In this latter single-mode regime, {the gap persists because the physical current $-pT\partial_x\ln(p/p^*)$ becomes roughly symmetric about the barrier top, draining one well while filling the other, whereas the optimal transport flux $p\partial_x\lambda$ is drawn mainly from the overpopulated well. Finally, in the vicinity of the steady state $p\simeq p^*$,} the physical flow and optimal transport fields become globally proportional therefore saturating the bound, {see also Fig. \ref{bit_erasure_fields} for a spatial comparison of the physical and optimal transport flows.}

\begin{figure}[ht]
    \centering
    \includegraphics[width=\columnwidth]{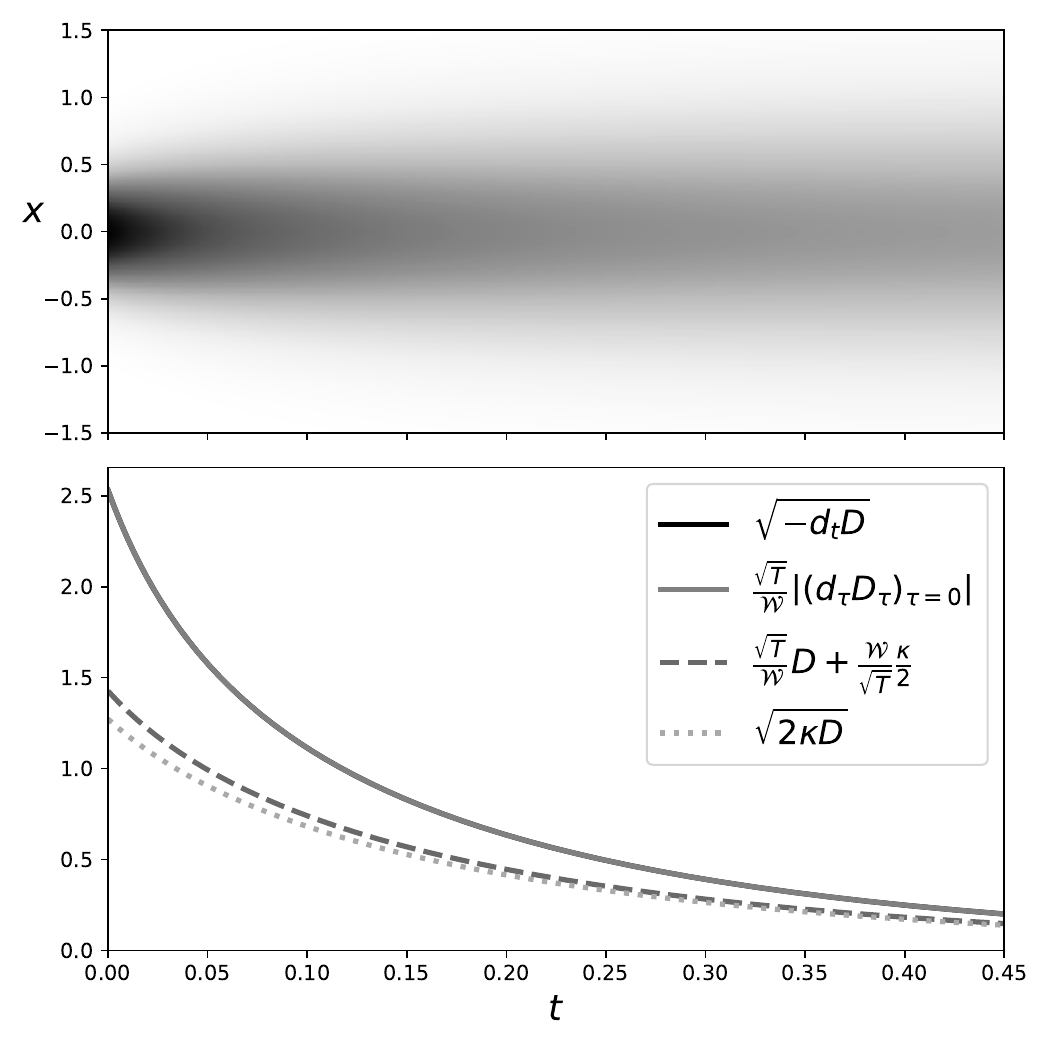}
    \caption{Thermal expansion in a harmonic potential. The parameters are set to $T=1$, $\kappa=2$, and initial stiffness $\gamma=10$. (Top) Symmetric expansion of the Gaussian probability density $p(x,t)$ toward the broader steady state. (Bottom) The lower bound of Eq. \eqref{general bound} (gray line) saturates at all times the instantaneous relaxation speed $\sqrt{-d_t D}$ (black line, hidden under the gray line). Here the bounds from the HWI [Eq. \eqref{HWI}] (gray dashed line) and LSI [Eq. \eqref{LogSobolev}] (gray dotted line) are non trivial.}
    \label{Gauss_expanding}
\end{figure}

\paragraph*{Saturation in a Gaussian expansion.} 

Consider a thermal expansion in a harmonic potential, $\psi(x, t\geq 0) = \frac{1}{2}\kappa x^2$. The system is initially prepared in thermal equilibrium inside a more confining potential $\psi(x, t< 0) = \frac{1}{2}\gamma x^2$ with $\gamma > \kappa$. At time $t=0$, the trap is instantaneously {changed $\gamma \to \kappa$} and the ensemble expands symmetrically toward a broader Gaussian steady state. 
The probability density $p(x,t)$ remains perfectly Gaussian at all times during the relaxation process, and the optimal transport map is linear. This case can be treated analytically, see the SM.
As shown in Fig. \ref{Gauss_expanding}, while the fundamental bound Eq. \eqref{general bound} is saturated, because the expansion changes the entropy of the system it holds strictly $\| \partial_x^2\lambda_\tau \| ^2 > 0$, therefore the HWI inequality Eq. \eqref{HWI} is not saturated. The LSI Eq. \eqref{LogSobolev} is looser than HWI as the corresponding Talagrand inequality $D \geq \kappa\mathcal{W}^2 / (2T) $ saturates only for pure translations.

{The reverse quench $\kappa\to\gamma$, which compresses the density along the same geodesic, is also Gaussian and saturates Eq.~\eqref{general bound}. For a harmonic potential, the HWI fully includes the energetic curvature in Eq.~\eqref{2nd derivative optimal transport} and neglects only the entropic term, which enters as an integrand 
$p_\tau(1-\tau) \|\partial_x^2\lambda_\tau\|^2$ in Eq.~\eqref{integral relation}.
At the quench, the two processes traverse the same geodesic in opposite directions, so this weighted integral is larger for the expansion, which starts at the narrower density where $\|\partial_x^2\lambda_\tau\|^2$ is largest.
Moreover, the exact slope $-(d_\tau D_\tau)_{\tau=0}=\int_0^1 d\tau\,d^2_\tau D_\tau$, which follows from $(d_\tau D_\tau)_{\tau=1}=0$, contains the unweighted integrand $p_\tau \|\partial_x^2\lambda_\tau\|^2$ which gives the same integral for both directions, plus an energetic term proportional to the final stiffness which is smaller for the expansion. The relative gap of the HWI is therefore larger for the expansion at the quench, and the SM shows that this ordering holds at all times $t$.}

\section{Discussion.}
In conclusion, functional inequalities in the Otto-Villani HWI framework provide a geometric perspective on the instantaneous relaxation speed of continuous systems. While the classical HWI inequality reduces the impact of the potential landscape to its global minimum convexity, the intermediate bound captures local details of the free energy landscape through a projection of the physical flow onto the optimal transport geodesic, and its tightness typically stays finite also in non-convex potentials.

While the Benamou--Brenier formulation of optimal transport has been fruitfully employed in the stochastic thermodynamics literature to derive thermodynamic speed limits, the Otto--Villani HWI framework appears to have remained unexplored in this context. Its lower bounds on the dissipation rate {are expressed in terms of the instantaneous density and the steady-state density,} and are thus complementary to existing thermodynamic speed limits~\cite{aurell2012refined,shiraishi2019information,dechant2019thermodynamic,nakazato2021geometrical,van2023thermodynamic}{, which are expressed in terms of the density at two different times along the process.}
{In particular, in the absence of nonconservative forces, such speed limits either vanish or they reduce to the definition of the entropy
production rate as the two times approach each other, see the SM, whereas Eq. \eqref{general bound} is a
nontrivial bound on the instantaneous relaxation speed, obtained by comparing
the direction of the physical flow with the geodesic toward the steady state.}

{Finally we note that the general bound Eq.~\eqref{general bound} is not meant here as a computational shortcut, since the relaxation speed is directly accessible from the instantaneous density if the steady state is known, while the bound requires solving an optimal transport problem. Its value is rather in the geometric intuition it offers, as it measures how far the physical relaxation deviates from the direct transport route toward the steady state, and it identifies the majorization steps that make the HWI and LSI inequalities loose in non-convex landscapes.}

\section*{Acknowledgements} I thank Sosuke Ito for fruitful discussions.


\bibliography{bibliography}

@PREAMBLE{
 "\providecommand{\noopsort}[1]{}" 
 # "\providecommand{\singleletter}[1]{#1}%" 
}

@book{karatzas2014brownian,
  title={Brownian motion and stochastic calculus},
  author={Karatzas, Ioannis and Shreve, Steven},
  year={2014},
  publisher={springer}
}

@article{ProesmansFinite2020,
  title = {Finite-Time Landauer Principle},
  author = {Proesmans, Karel and Ehrich, Jannik and Bechhoefer, John},
  journal = {Phys. Rev. Lett.},
  volume = {125},
  issue = {10},
  pages = {100602},
  numpages = {6},
  year = {2020},
  month = {Sep},
  publisher = {American Physical Society}
  }

@article{lu2017nonequilibrium,
  title={Nonequilibrium thermodynamics of the Markovian Mpemba effect and its inverse},
  author={Lu, Zhiyue and Raz, Oren},
  journal={Proceedings of the National Academy of Sciences},
  volume={114},
  number={20},
  pages={5083--5088},
  year={2017},
  publisher={National Academy of Sciences}
}

@article{teza2026speedups,
  title={Speedups in nonequilibrium thermal relaxation: Mpemba and related effects},
  author={Teza, Gianluca and Bechhoefer, John and Lasanta, Antonio and Raz, Oren and Vucelja, Marija},
  journal={Physics Reports},
  volume={1164},
  pages={1--97},
  year={2026},
  publisher={Elsevier}
}

@article{hatano2001steady,
  title={Steady-state thermodynamics of Langevin systems},
  author={Hatano, Takahiro and Sasa, Shin-ichi},
  journal={Physical review letters},
  volume={86},
  number={16},
  pages={3463},
  year={2001},
  publisher={APS}
}

@article{bao2025universal,
  title={Universal trade-off between irreversibility and intrinsic timescale in thermal relaxation with applications to thermodynamic inference},
  author={Bao, Ruicheng and Du, Chaoqun and Cao, Zhiyu and Hou, Zhonghuai},
  journal={Physical Review E},
  volume={112},
  number={4},
  pages={044134},
  year={2025},
  publisher={APS}
}

@book{amari2016information,
  title={Information geometry and its applications},
  author={Amari, Shun-ichi},
  year={2016},
  publisher={Springer}
}

@article{kolchinsky2026generalized,
  title={Generalized free energy and excess/housekeeping decomposition in nonequilibrium systems: From large deviations to thermodynamic speed limits},
  author={Kolchinsky, Artemy and Dechant, Andreas and Yoshimura, Kohei and Ito, Sosuke},
  journal={Physical Review Research},
  volume={8},
  number={2},
  pages={023025},
  year={2026},
  publisher={APS}
}

@article{ito2022information,
  title={Information geometry, trade-off relations, and generalized Glansdorff--Prigogine criterion for stability},
  author={Ito, Sosuke},
  journal={Journal of Physics A: Mathematical and Theoretical},
  volume={55},
  number={5},
  pages={054001},
  year={2022},
  publisher={IOP Publishing}
}

@article{maes2015revisiting,
  title={Revisiting the Glansdorff--Prigogine criterion for stability within irreversible thermodynamics},
  author={Maes, Christian and Neto{\v{c}}n{\`y}, Karel},
  journal={Journal of Statistical Physics},
  volume={159},
  number={6},
  pages={1286--1299},
  year={2015},
  publisher={Springer}
}

@article{glansdorff1974thermodynamic,
  title={The thermodynamic stability theory of non-equilibrium states},
  author={Glansdorff, Paul and Nicolis, G and Prigogine, I},
  journal={Proceedings of the National Academy of Sciences},
  volume={71},
  number={1},
  pages={197--199},
  year={1974}
}

@article{van2023thermodynamic,
	title={Thermodynamic unification of optimal transport: Thermodynamic uncertainty relation, minimum dissipation, and thermodynamic speed limits},
	author={Van Vu, Tan and Saito, Keiji},
	journal={Physical Review X},
	volume={13},
	number={1},
	pages={011013},
	year={2023},
	publisher={APS}
}

@article{lan2012energy,
  title={The energy--speed--accuracy trade-off in sensory adaptation},
  author={Lan, Ganhui and Sartori, Pablo and Neumann, Silke and Sourjik, Victor and Tu, Yuhai},
  journal={Nature physics},
  volume={8},
  number={5},
  pages={422--428},
  year={2012},
  publisher={Nature Publishing Group UK London}
}

@article{ito2015maxwell,
  title={Maxwell’s demon in biochemical signal transduction with feedback loop},
  author={Ito, Sosuke and Sagawa, Takahiro},
  journal={Nature communications},
  volume={6},
  number={1},
  pages={7498},
  year={2015},
  publisher={Nature Publishing Group UK London}
}

@article{edelson2025context,
  title={Context dependent adaptation in a neural computation},
  author={Edelson, Charles J and Setayeshgar, Sima and Bialek, William and van Steveninck, Rob R},
  journal={arXiv preprint arXiv:2509.01760},
  year={2025}
}

@article{erbar2015equivalence,
  title={On the equivalence of the entropic curvature-dimension condition and Bochner’s inequality on metric measure spaces},
  author={Erbar, Matthias and Kuwada, Kazumasa and Sturm, Karl-Theodor},
  journal={Inventiones mathematicae},
  volume={201},
  number={3},
  pages={993--1071},
  year={2015},
  publisher={Springer}
}

@book{peyre2019computational,
  title={Computational optimal transport: With applications to data science},
  author={Peyr{\'e}, Gabriel and Cuturi, Marco},
  year={2019},
  publisher={Now Foundations and Trends}
}

@article{dembo2002information,
  title={Information theoretic inequalities},
  author={Dembo, Amir and Cover, Thomas M and Thomas, Joy A},
  journal={IEEE Transactions on Information theory},
  volume={37},
  number={6},
  pages={1501--1518},
  year={2002},
  publisher={IEEE}
}

@article{ito2020stochastic,
  title={Stochastic time evolution, information geometry, and the Cram{\'e}r-Rao bound},
  author={Ito, Sosuke and Dechant, Andreas},
  journal={Physical Review X},
  volume={10},
  number={2},
  pages={021056},
  year={2020},
  publisher={APS}
}

@article{auconi2026information,
  title={Information-Geometric Signatures of Nonconservative Driving},
  author={Auconi, Andrea and Ito, Sosuke},
  journal={arXiv preprint arXiv:2605.03757},
  year={2026}
}

@article{ito2024geometric,
  title={Geometric thermodynamics for the Fokker--Planck equation: stochastic thermodynamic links between information geometry and optimal transport},
  author={Ito, Sosuke},
  journal={Information geometry},
  volume={7},
  number={Suppl 1},
  pages={441--483},
  year={2024},
  publisher={Springer}
}

@article{benamou2000computational,
	title={A computational fluid mechanics solution to the Monge-Kantorovich mass transfer problem},
	author={Benamou, Jean-David and Brenier, Yann},
	journal={Numerische Mathematik},
	volume={84},
	number={3},
	pages={375--393},
	year={2000},
	publisher={Springer-Verlag Berlin/Heidelberg}
}

@article{dechant2019thermodynamic,
  title={Thermodynamic interpretation of Wasserstein distance},
  author={Dechant, Andreas and Sakurai, Yohei},
  journal={arXiv preprint arXiv:1912.08405},
  year={2019}
}

@article{aurell2012refined,
  title={Refined second law of thermodynamics for fast random processes},
  author={Aurell, Erik and Gaw{\c{e}}dzki, Krzysztof and Mej{\'\i}a-Monasterio, Carlos and Mohayaee, Roya and Muratore-Ginanneschi, Paolo},
  journal={Journal of statistical physics},
  volume={147},
  number={3},
  pages={487--505},
  year={2012},
  publisher={Springer}
}

@article{karatzas2022trajectorial,
  title={A Trajectorial Approach to the Gradient Flow Properties of Langevin--Smoluchowski Diffusions},
  author={Karatzas, Ioannis and Schachermayer, Walter and Tschiderer, Bertram},
  journal={Theory of Probability \& Its Applications},
  volume={66},
  number={4},
  pages={668--707},
  year={2022},
  publisher={SIAM}
}

@article{gentil2020entropic,
  title={An entropic interpolation proof of the HWI inequality},
  author={Gentil, Ivan and L{\'e}onard, Christian and Ripani, Luigia and Tamanini, Luca},
  journal={Stochastic Processes and their Applications},
  volume={130},
  number={2},
  pages={907--923},
  year={2020},
  publisher={Elsevier}
}

@article{lott2009ricci,
  title={Ricci curvature for metric-measure spaces via optimal transport},
  author={Lott, John and Villani, C{\'e}dric},
  journal={Annals of Mathematics},
  pages={903--991},
  year={2009},
  publisher={JSTOR}
}

@article{dechant2022geometric,
  title={Geometric decomposition of entropy production in out-of-equilibrium systems},
  author={Dechant, Andreas and Sasa, Shin-ichi and Ito, Sosuke},
  journal={Physical Review Research},
  volume={4},
  number={1},
  pages={L012034},
  year={2022},
  publisher={APS}
}

@article{nakazato2021geometrical,
  title={Geometrical aspects of entropy production in stochastic thermodynamics based on Wasserstein distance},
  author={Nakazato, Muka and Ito, Sosuke},
  journal={Physical Review Research},
  volume={3},
  number={4},
  pages={043093},
  year={2021},
  publisher={APS}
}

@article{seifert2012stochastic,
  title={Stochastic thermodynamics, fluctuation theorems and molecular machines},
  author={Seifert, Udo},
  journal={Reports on progress in physics},
  volume={75},
  number={12},
  pages={126001},
  year={2012},
  publisher={IOP Publishing}
}

@incollection{risken1989fokker,
  title={Fokker-planck equation},
  author={Risken, Hannes},
  booktitle={The Fokker-Planck equation: methods of solution and applications},
  pages={63--95},
  year={1989},
  publisher={Springer}
}

@book{villani2009optimal,
	title={Optimal transport: old and new},
	author={Villani, C{\'e}dric and others},
	volume={338},
	year={2009},
	publisher={Springer}
}

@article{otto2000generalization,
	title={Generalization of an inequality by Talagrand and links with the logarithmic Sobolev inequality},
	author={Otto, Felix and Villani, C{\'e}dric},
	journal={Journal of Functional Analysis},
	volume={173},
	number={2},
	pages={361--400},
	year={2000},
	publisher={Elsevier}
}

@article{auconi2025nonequilibrium,
	title={Nonequilibrium relaxation inequality on short timescales},
	author={Auconi, Andrea},
	journal={Physical Review Letters},
	volume={134},
	number={8},
	pages={087104},
	year={2025},
	publisher={APS}
}

@book{ekeland1999convex,
  title={Convex analysis and variational problems},
  author={Ekeland, Ivar and Temam, Roger},
  year={1999},
  publisher={SIAM}
}

@article{shiraishi2019information,
  title={Information-theoretical bound of the irreversibility in thermal relaxation processes},
  author={Shiraishi, Naoto and Saito, Keiji},
  journal={Physical review letters},
  volume={123},
  number={11},
  pages={110603},
  year={2019},
  publisher={APS}
}

@article{berut2012experimental,
  title={Experimental verification of Landauer’s principle linking information and thermodynamics},
  author={B{\'e}rut, Antoine and Arakelyan, Artak and Petrosyan, Artyom and Ciliberto, Sergio and Dillenschneider, Raoul and Lutz, Eric},
  journal={Nature},
  volume={483},
  number={7388},
  pages={187--189},
  year={2012},
  publisher={Nature Publishing Group UK London}
}

@article{lapolla2020faster,
  title={Faster uphill relaxation in thermodynamically equidistant temperature quenches},
  author={Lapolla, Alessio and Godec, Alja{\v{z}}},
  journal={Physical Review Letters},
  volume={125},
  number={11},
  pages={110602},
  year={2020},
  publisher={APS}
}

@article{meibohm2021relaxation,
  title={Relaxation-speed crossover in anharmonic potentials},
  author={Meibohm, Jan and Forastiere, Danilo and Adeleke-Larodo, Tunrayo and Proesmans, Karel},
  journal={Physical Review E},
  volume={104},
  number={3},
  pages={L032105},
  year={2021},
  publisher={APS}
}

@article{ibanez2024heating,
  title={Heating and cooling are fundamentally asymmetric and evolve along distinct pathways},
  author={Ib{\'a}{\~n}ez, Miguel and Dieball, Cai and Lasanta, Antonio and Godec, Alja{\v{z}} and Rica, Ra{\'u}l A},
  journal={Nature Physics},
  volume={20},
  number={1},
  pages={135--141},
  year={2024},
  publisher={Nature Publishing Group UK London}
}

@article{PhysRevLett.121.070601,
  title = {Speed Limit for Classical Stochastic Processes},
  author = {Shiraishi, Naoto and Funo, Ken and Saito, Keiji},
  journal = {Phys. Rev. Lett.},
  volume = {121},
  issue = {7},
  pages = {070601},
  numpages = {6},
  year = {2018},
  month = {Aug},
  publisher = {American Physical Society}
}

@MISC{Note1,note="This exact form is not explicitly written in the original Ref. \cite{otto2000generalization}, but it is an immediate intermediate bound within the HWI inequality derivation therein."}

@article{jordan1998variational,
  title={The variational formulation of the Fokker--Planck equation},
  author={Jordan, Richard and Kinderlehrer, David and Otto, Felix},
  journal={SIAM journal on mathematical analysis},
  volume={29},
  number={1},
  pages={1--17},
  year={1998},
  publisher={SIAM}
}

@book{villani2021topics,
  title={Topics in optimal transportation},
  author={Villani, C{\'e}dric},
  volume={58},
  year={2021},
  publisher={American Mathematical Soc.}
}

@book{kubo2012statistical,
  title={Statistical physics II: nonequilibrium statistical mechanics},
  author={Kubo, Ryogo and Toda, Morikazu and Hashitsume, Natsuki},
  volume={31},
  year={2012},
  publisher={Springer Science \& Business Media}
}

@article{lee2022speed,
  title={Speed limit for a highly irreversible process and tight finite-time Landauer’s bound},
  author={Lee, Jae Sung and Lee, Sangyun and Kwon, Hyukjoon and Park, Hyunggyu},
  journal={Physical review letters},
  volume={129},
  number={12},
  pages={120603},
  year={2022},
  publisher={APS}
}

@article{dechant2018entropic,
  title={Entropic bounds on currents in Langevin systems},
  author={Dechant, Andreas and Sasa, Shin-ichi},
  journal={Physical Review E},
  volume={97},
  number={6},
  pages={062101},
  year={2018},
  publisher={APS}
}

@book{gaspard2022statistical,
  title={The statistical mechanics of irreversible phenomena},
  author={Gaspard, Pierre},
  year={2022},
  publisher={Cambridge University Press}
}

@article{zhong2024beyond,
  title={Beyond linear response: Equivalence between thermodynamic geometry and optimal transport},
  author={Zhong, Adrianne and DeWeese, Michael R},
  journal={Physical Review Letters},
  volume={133},
  number={5},
  pages={057102},
  year={2024},
  publisher={APS}
}

@article{shiraishi2024wasserstein,
  title={Wasserstein distance in speed limit inequalities for Markov jump processes},
  author={Shiraishi, Naoto},
  journal={Journal of Statistical Mechanics: Theory and Experiment},
  volume={2024},
  number={7},
  pages={074003},
  year={2024},
  publisher={IOP Publishing}
}

@article{sabbagh2024wasserstein,
  title={Wasserstein speed limits for Langevin systems},
  author={Sabbagh, Ralph and Movilla Miangolarra, Olga and Georgiou, Tryphon T},
  journal={Physical Review Research},
  volume={6},
  number={3},
  pages={033308},
  year={2024},
  publisher={APS}
}

@inproceedings{rioul2017optimal,
  title={Optimal transportation to the entropy-power inequality},
  author={Rioul, Olivier},
  booktitle={2017 Information Theory and Applications Workshop (ITA)},
  pages={1--5},
  year={2017},
  organization={IEEE}
}

@article{oikawa2025experimentally,
  title={Experimentally achieving minimal dissipation via thermodynamically optimal transport},
  author={Oikawa, Shingo and Nakayama, Yohei and Ito, Sosuke and Sagawa, Takahiro and Toyabe, Shoichi},
  journal={Nature Communications},
  volume={16},
  number={1},
  pages={10424},
  year={2025},
  publisher={Nature Publishing Group UK London}
}

@article{esposito2011second,
  title={Second law and Landauer principle far from equilibrium},
  author={Esposito, Massimiliano and Van den Broeck, Christian},
  journal={EPL (Europhysics Letters)},
  volume={95},
  number={4},
  pages={40004},
  year={2011}
}

@article{van2010three,
  title={Three faces of the second law. II. Fokker-Planck formulation},
  author={Van den Broeck, Christian and Esposito, Massimiliano},
  journal={Physical Review E—Statistical, Nonlinear, and Soft Matter Physics},
  volume={82},
  number={1},
  pages={011144},
  year={2010},
  publisher={APS}
}

\end{document}


\preprint{APS/123-QED}

\title{Supplementary Materials for the manuscript ``Thermodynamic Bounds from Otto--Villani Functional Inequalities"}




\maketitle

\subsection{Fokker-Planck equation and Logarithmic Sobolev inequality}
\label{sec:FP_LSI}

Consider the overdamped Fokker-Planck equation, that is Eq. (2) in the main text,
\begin{equation}\label{FP}
    \partial_t p = \boldsymbol{\nabla}\cdot \left[ p \boldsymbol{\nabla}(\psi + T\ln p ) \right] = T \boldsymbol{\nabla} \cdot \left[ p \boldsymbol{\nabla} \ln \left( \frac{p}{p^*} \right) \right],
\end{equation}
where in the last step the invariant density $p^* = (1/Z) \exp(-\psi/T)$ was considered. 
The Kullback-Leibler (KL) divergence \cite{amari2016information} from the steady state is defined as $D \equiv D[p\|p^*] = \int d\boldsymbol{x} \, p \ln(p/p^*)$. Its first time derivative along the physical dynamics Eq. \eqref{FP} is computed as
\begin{multline}\label{first derivative phys}
d_t D = \int d\boldsymbol{x} \, (\partial_t p) \ln \left( \frac{p}{p^*} \right) \\
=T \int d\boldsymbol{x} \, \ln \left( \frac{p}{p^*} \right) \boldsymbol{\nabla} \cdot \left[ p \boldsymbol{\nabla} \ln \left( \frac{p}{p^*} \right) \right] \\
= -T \int d\boldsymbol{x} \, p \left\| \boldsymbol{\nabla} \ln \left( \frac{p}{p^*} \right) \right\|^2 \le 0,
\end{multline}
where probability conservation was considered, $\int d\boldsymbol{x} \, \partial_t p = 0$, and integration by parts was performed assuming $p \ln (p/p^*) \boldsymbol{\nabla} \ln (p/p^*)$ vanishes fast enough at infinity.
The negativity $d_t D \leq 0$ guarantees the stability of the steady state.

Let us define $h \equiv \ln(p/p^*)$, whose time derivative reads
\begin{multline} \label{d_t h}
\partial_t h = T \nabla^2 h +T \boldsymbol{\nabla} \ln p \cdot \boldsymbol{\nabla} h
\\
= T \nabla^2 h + T \|\boldsymbol{\nabla} h\|^2 - \boldsymbol{\nabla} \psi \cdot \boldsymbol{\nabla} h.
\end{multline}
Eq. \eqref{first derivative phys} can be rewritten in terms of $h$ as $d_t D = -T \int d\boldsymbol{x} \, p \|\boldsymbol{\nabla} h\|^2$.
The second time derivative is composed of two terms,
\begin{multline}\label{pre d^2_t D}
-\frac{1}{T} d_t^2 D \\= \int d\boldsymbol{x} \, (\partial_t p) \|\boldsymbol{\nabla} h\|^2 +2 \int d\boldsymbol{x} \, p \boldsymbol{\nabla} h \cdot \boldsymbol{\nabla} \partial_t h.
\end{multline}
The first term is evaluated as
\begin{multline} \label{term one}
\int d\boldsymbol{x} \, (\partial_t p) \|\boldsymbol{\nabla} h\|^2 = 
T \int d\boldsymbol{x} \,  \|\boldsymbol{\nabla} h\|^2  \boldsymbol{\nabla} \cdot (p \boldsymbol{\nabla} h) \\
= -T \int d\boldsymbol{x} \, p \boldsymbol{\nabla} h \cdot \boldsymbol{\nabla} (\|\boldsymbol{\nabla} h\|^2) \\
= -2T \int d\boldsymbol{x} \, p \boldsymbol{\nabla} h \cdot \boldsymbol{\nabla} \boldsymbol{\nabla} h \cdot \boldsymbol{\nabla} h,
\end{multline}
where integration by parts was performed assuming $p \|\boldsymbol{\nabla} h\|^2 \boldsymbol{\nabla} h$ vanishes fast enough at infinity.
The second term is evaluated from Eq. \eqref{d_t h} as
\begin{multline}\label{term two}
 \int d\boldsymbol{x} \, p \boldsymbol{\nabla} h \cdot \boldsymbol{\nabla} (\partial_t h) \\
 =\int d\boldsymbol{x} \, p \boldsymbol{\nabla} h \cdot \boldsymbol{\nabla} \left( T \nabla^2 h + T \|\boldsymbol{\nabla} h\|^2 - \boldsymbol{\nabla} \psi \cdot \boldsymbol{\nabla} h \right)\\
= T \int d\boldsymbol{x} \, p \boldsymbol{\nabla} h \cdot \boldsymbol{\nabla} (\nabla^2 h) +2T \int d\boldsymbol{x} \, p \boldsymbol{\nabla} h \cdot \boldsymbol{\nabla} \boldsymbol{\nabla} h \cdot \boldsymbol{\nabla} h \\
- \int d\boldsymbol{x} \, p \boldsymbol{\nabla} h \cdot \boldsymbol{\nabla} \boldsymbol{\nabla} \psi \cdot \boldsymbol{\nabla} h - \int d\boldsymbol{x} \, p \boldsymbol{\nabla} h \cdot \boldsymbol{\nabla} \boldsymbol{\nabla} h \cdot \boldsymbol{\nabla} \psi.
\end{multline}

Consider the vector calculus identity
\begin{multline}
\boldsymbol{\nabla} \cdot \left( p \boldsymbol{\nabla}\boldsymbol{\nabla} h \cdot \boldsymbol{\nabla} h \right) = p \|\boldsymbol{\nabla}\boldsymbol{\nabla} h\|^2 \\+ p \boldsymbol{\nabla} h \cdot \boldsymbol{\nabla} \nabla^2 h + \boldsymbol{\nabla} p \cdot \boldsymbol{\nabla}\boldsymbol{\nabla} h \cdot \boldsymbol{\nabla} h,
\end{multline}
where the square of the Frobenius norm is defined as
\begin{equation}
\|\boldsymbol{\nabla} \boldsymbol{\nabla} h\|^2 = \boldsymbol{\nabla} \boldsymbol{\nabla} h \cdot \boldsymbol{\nabla} \boldsymbol{\nabla} h
\equiv \sum_{i,j=1}^n \left( \frac{\partial^2 h}{\partial x_i \partial x_j} \right)^2.
\end{equation}
Assuming $p (\boldsymbol{\nabla}\boldsymbol{\nabla} h) \cdot \boldsymbol{\nabla} h$ to vanish fast enough at infinity, integrating this identity over space yields
\begin{multline}\label{Frobenius by part}
\int d\boldsymbol{x} \, p \|\boldsymbol{\nabla} \boldsymbol{\nabla} h\|^2 \\= - \int d\boldsymbol{x} \, p \boldsymbol{\nabla} h \cdot \boldsymbol{\nabla} \nabla^2 h 
- \int d\boldsymbol{x} \, \boldsymbol{\nabla} p \cdot \boldsymbol{\nabla} \boldsymbol{\nabla} h \cdot \boldsymbol{\nabla} h \\
= - \int d\boldsymbol{x} \, p \boldsymbol{\nabla} h \cdot \boldsymbol{\nabla} \nabla^2 h 
-  \int d\boldsymbol{x} \, p \boldsymbol{\nabla} h \cdot \boldsymbol{\nabla} \boldsymbol{\nabla} h \cdot \boldsymbol{\nabla} h \\+ \frac{1}{T} \int d\boldsymbol{x} \, p \boldsymbol{\nabla} \psi \cdot \boldsymbol{\nabla} \boldsymbol{\nabla} h \cdot \boldsymbol{\nabla} h.
\end{multline}

From Eqs. \eqref{pre d^2_t D}-\eqref{Frobenius by part}, and considering the Hessian symmetry $\boldsymbol{\nabla} \psi \cdot \boldsymbol{\nabla} \boldsymbol{\nabla} h \cdot \boldsymbol{\nabla} h=\boldsymbol{\nabla} h \cdot \boldsymbol{\nabla} \boldsymbol{\nabla} h \cdot \boldsymbol{\nabla} \psi$,
one obtains
\begin{equation} \label{second derivative phys}
d_t^2 D = 2T \int d\boldsymbol{x} \, p \left( T \|\boldsymbol{\nabla} \boldsymbol{\nabla} h\|^2 + \boldsymbol{\nabla} h \cdot \boldsymbol{\nabla} \boldsymbol{\nabla} \psi \cdot \boldsymbol{\nabla} h \right),
\end{equation}
which is the standard Bakry-\'Emery form of the second derivative \cite{villani2021topics}.

Consider the minimum convexity of the potential defined via the Loewner order as $\kappa \equiv \inf_x \lambda_{\text{min}}(\boldsymbol{\nabla} \boldsymbol{\nabla} \psi)$. From the corresponding global majorization $\boldsymbol{\nabla} h \cdot \boldsymbol{\nabla} \boldsymbol{\nabla} \psi \cdot \boldsymbol{\nabla} h \ge \kappa \|\boldsymbol{\nabla} h\|^2$ and the nonnegativity of the Frobenius norm $\|\boldsymbol{\nabla} \boldsymbol{\nabla} h\|^2 \ge 0$, the second derivative in Eq. \eqref{second derivative phys} is bounded by
\begin{equation}
d_t^2 D \ge 2T\kappa \int d\boldsymbol{x} \, p \|\boldsymbol{\nabla} h\|^2
= -2\kappa d_t D.
\end{equation}
Integrating this inequality over time from $t=0$ to $t=\infty$, and considering the boundary conditions $\lim_{t \to \infty} d_t D = \lim_{t \to \infty} D = 0$ given by the assumed asymptotic convergence, one obtains
\begin{equation}
-d_t D \ge 2\kappa D,
\end{equation}
that is the Logarithmic Sobolev inequality (LSI).
When $\kappa > 0$, the LSI implies the exponential convergence to the steady state as $D[p(t) \| p^*] \le D[p(0) \| p^*]\exp(-2\kappa t)$.

\subsection{Relation to physical observables}
\label{sec:Relation to observables}

The relaxation speed of the free energy $d_t D$ is related to that of any observable $\langle f \rangle \equiv \int d\boldsymbol{x}\, p f(\boldsymbol{x})$ by the below inequality \cite{dechant2018entropic,ito2020stochastic},
\begin{equation}
    -d_t D \geq \frac{( d_t \langle f \rangle )^2}{T\langle \| \boldsymbol{\nabla}f \|^2 \rangle} .
\end{equation}
This follows from integration by parts and the Cauchy-Schwarz inequality as
\begin{multline}
    ( d_t \langle f \rangle )^2 =\left( \int d\boldsymbol{x} \, f \partial_t p \right)^2 \\
    = \left( \int d\boldsymbol{x} \, f \boldsymbol{\nabla}\cdot \left[ p \boldsymbol{\nabla} (\psi + T\ln p )  \right] \right)^2 \\
    =  \left( \int d\boldsymbol{x} \, p \boldsymbol{\nabla}f \cdot \left[ \boldsymbol{\nabla} (\psi + T\ln p )  \right] \right)^2 \\
    \leq \left( \int d\boldsymbol{x} \, p \| \boldsymbol{\nabla}f \| \, \| \boldsymbol{\nabla} (\psi + T\ln p ) \| \right)^2 \\
    \leq \left( \int d\boldsymbol{x} \, p \| \boldsymbol{\nabla}f \|^2 \right) \left( \int d\boldsymbol{x} \, p \| \boldsymbol{\nabla} (\psi + T\ln p ) \|^2 \right) \\
     = \langle \| \boldsymbol{\nabla}f \|^2 \rangle (-T d_t D) .
\end{multline}

\subsection{Review of Optimal Transport Theory}
\label{sec:Review_Optimal_Transport}

Consider a fluid with density $p_\tau$ and velocity field $\boldsymbol{v}_\tau$, evolving in the fictitious transport time $\tau$ from $\tau=0$ to $\tau=1$, where as in the main text the subscript $\tau$ denotes the dependence of a spatial field on the transport time. 
The optimal transport problem \cite{benamou2000computational,villani2009optimal} is the minimization of the kinetic action $ \mathcal{A}\equiv  \mathcal{A}[p_\tau, \boldsymbol{v}_\tau]$ defined by
\begin{equation}\label{def A}
    \mathcal{A} \equiv \frac{1}{2} \int_0^1 d\tau \int d\boldsymbol{x} \, p_\tau \|\boldsymbol{v}_\tau\|^2 ,
\end{equation}
subject to the continuity equation $\partial_\tau p_\tau = - \boldsymbol{\nabla} \cdot (p_\tau \boldsymbol{v}_\tau)$, and to given initial and final densities respectively $p_0 = p$ and $p_1 = q$.

The continuity equation constraint is accounted for by introducing the Lagrangian $\mathcal{L} \equiv \mathcal{L}[p_\tau, \boldsymbol{v}_\tau, \lambda_\tau]$ as
\begin{equation}\label{Lagrangian}
    \mathcal{L} \equiv \mathcal{A} + \int_0^1 d\tau \int d\boldsymbol{x} \, \lambda_\tau \left[ \partial_\tau p_\tau + \boldsymbol{\nabla} \cdot (p_\tau \boldsymbol{v}_\tau) \right] ,
\end{equation} 
in terms of a smooth multiplier field $\lambda_\tau$, and by imposing the supremum
\begin{equation}\label{inf sup}
    \inf_{p_\tau>0,\boldsymbol{v}_\tau} \sup_{\lambda_\tau} \mathcal{L},
\end{equation}
as this ensures a positive diverging contribution when $\partial_\tau p_\tau \neq - \boldsymbol{\nabla} \cdot (p_\tau \boldsymbol{v}_\tau)$, which then gets discarded by the infimum coming from the original minimization problem.
Note that $p_\tau > 0$ has been assumed to avoid vacuum states.

Integrating by parts the continuity equation terms in Eq. \eqref{Lagrangian}, in particular $\partial_\tau p_\tau$ with respect to time inverting the order of integration, and $\boldsymbol{\nabla} \cdot (p_\tau \boldsymbol{v}_\tau)$ with respect to space assuming $\lambda_\tau p_\tau \|\boldsymbol{v}_\tau\|$ to decay fast enough at infinity, the Lagrangian becomes
\begin{multline}
    \mathcal{L} = \int d\boldsymbol{x}  \Big[\left[\lambda_\tau p_\tau\right]_0^1 \\ + \int_0^1 d\tau \, p_\tau \left( \frac{1}{2} \|\boldsymbol{v}_\tau\|^2 - \partial_\tau \lambda_\tau - \boldsymbol{v}_\tau \cdot \boldsymbol{\nabla} \lambda_\tau \right)  \Big] ,
\end{multline}
where $\left[\lambda_\tau p_\tau\right]_0^1 \equiv \lambda_1 p_1 -\lambda_0 p_0$, and the fluid variables $p_\tau$ and $\boldsymbol{v}_\tau$ are now free of differential constraints.

Following Benamou and Brenier \cite{benamou2000computational}, let us consider the momentum density $\boldsymbol{m}_\tau \equiv p_\tau \boldsymbol{v}_\tau$, and denote the Lagrangian in the $(p_\tau, \boldsymbol{m}_\tau)$ coordinates as $\widetilde{\mathcal{L}}[p_\tau, \boldsymbol{m}_\tau, \lambda_\tau] = \mathcal{L}[p_\tau, \boldsymbol{v}_\tau, \lambda_\tau]|_{\boldsymbol{v}_\tau=\boldsymbol{m}_\tau/p_\tau}$.
The transformed Lagrangian is jointly convex in $(p_\tau, \boldsymbol{m}_\tau)$ and linear in $\lambda_\tau$. Therefore it satisfies the conditions of Sion's Minimax Theorem \cite{ekeland1999convex}, that allows to exchange the optimization order,
\begin{equation}\label{minimax 1}
    \inf_{p_\tau>0,\boldsymbol{m}_\tau} \sup_{\lambda_\tau} \widetilde{\mathcal{L}} 
    = \sup_{\lambda_\tau} \inf_{p_\tau>0, \boldsymbol{m}_\tau}\widetilde{\mathcal{L}} .
\end{equation}
Given the assumption $p_\tau > 0$, the mapping from $(p_\tau, \boldsymbol{v}_\tau)$ to $(p_\tau, \boldsymbol{m}_\tau)$ is invertible, therefore the optimization problem is invariant under this transformation,
\begin{equation}\label{minimax 2}
    \inf_{p_\tau>0,\boldsymbol{v}_\tau} \sup_{\lambda_\tau} \mathcal{L} = \inf_{p_\tau>0,\boldsymbol{m}_\tau} \sup_{\lambda_\tau} \widetilde{\mathcal{L}} .
\end{equation}
Then from Eqs. \eqref{minimax 1}-\eqref{minimax 2}, in terms of the original coordinates it holds
\begin{equation}
    \inf_{p_\tau>0,\boldsymbol{v}_\tau} \sup_{\lambda_\tau} \mathcal{L} = \sup_{\lambda_\tau} \inf_{p_\tau>0,\boldsymbol{v}_\tau} \mathcal{L} .
\end{equation}

The stationarity of the Lagrangian with respect to variations of the velocity field, ${\delta \mathcal{L}}/{\delta \boldsymbol{v}_\tau} = 0$, yields the optimality condition
\begin{equation}\label{v optimal}
     \boldsymbol{v}_\tau = \boldsymbol{\nabla} \lambda_\tau .
\end{equation}
Substituting this back into the Lagrangian yields 
\begin{multline}\label{sup inf argument}
    \sup_{\lambda_\tau} \inf_{p_\tau > 0, \boldsymbol{v}_\tau}\mathcal{L} =  \sup_{\lambda_\tau} \int d\boldsymbol{x} \Big[ \left[\lambda_\tau p_\tau\right]_0^1 \\
    + \int_0^1 d\tau \, \inf_{p_\tau > 0} p_\tau \left( -\frac{1}{2} \|\boldsymbol{\nabla} \lambda_\tau\|^2 - \partial_\tau \lambda_\tau \right) \Big] \\
    =  \sup_{\lambda_\tau} \left\{ \int d\boldsymbol{x} \, \left[\lambda_\tau p_\tau\right]_0^1 \, \bigg| \, \partial_\tau \lambda_\tau \leq -\frac{1}{2} \|\boldsymbol{\nabla} \lambda_\tau\|^2  \right\} ,
\end{multline}
where the last equality follows from recognizing that the infimum logic is such that configurations which admit somewhere $\partial_\tau \lambda_\tau > -\frac{1}{2} \|\boldsymbol{\nabla} \lambda_\tau\|^2 $ are correspondingly multiplied by a diverging $p_\tau\rightarrow\infty$, while contributions with $\partial_\tau \lambda_\tau < -\frac{1}{2} \|\boldsymbol{\nabla} \lambda_\tau\|^2 $ are multiplied by a vanishing $p_\tau\rightarrow 0$.
Note that $\inf_{p_\tau > 0, \boldsymbol{v}_\tau}$ does not affect $\left[\lambda_\tau p_\tau\right]_0^1$ since $p_0$ and $p_1$ are fixed conditions.

The uniform saturation case of the dynamic constraint $\partial_\tau \lambda_\tau \leq -\frac{1}{2} \|\boldsymbol{\nabla} \lambda_\tau\|^2$ is the Hamilton-Jacobi equation,
\begin{equation}\label{HJ}
    \partial_\tau \lambda_\tau = -\frac{1}{2} \|\boldsymbol{\nabla} \lambda_\tau\|^2 .
\end{equation}
Its solution is given by the Hopf-Lax formula,
\begin{equation}\label{Hopf-Lax}
    \lambda_\tau(\boldsymbol{x}) = \inf_{\boldsymbol{y}} \left\{ \lambda_0(\boldsymbol{y}) + \frac{\|\boldsymbol{x} - \boldsymbol{y}\|^2}{2\tau} \right\} ,
\end{equation}
as it can be verified by differentiation noting that the implicit derivative vanishes at the optimal minimizer $\boldsymbol{y^*}$,
\begin{equation}
    \frac{\partial}{\partial y_j} \left[ \lambda_0(\boldsymbol{y}) + \frac{\|\boldsymbol{x} - \boldsymbol{y}\|^2}{2\tau} \right] \, \bigg| \,_{\boldsymbol{y^*}} = 0.
\end{equation}

Let us now consider a multiplier trajectory $\mu_\tau$ different from the Hopf-Lax formula Eq. \eqref{Hopf-Lax}, but starting from the same configuration $\mu_0 = \lambda_0$.
Given a test path $\boldsymbol{z}(\tau) = \boldsymbol{y} + \tau(\boldsymbol{x} - \boldsymbol{y})$, with $\tau \in [0,1]$ and arbitrary $(\boldsymbol{x}, \boldsymbol{y})$, the corresponding total derivative of any admissible field $\mu_\tau(\boldsymbol{z}(\tau))$ reads 
\begin{multline}
	d_\tau \mu_\tau = \partial_\tau \mu_\tau +  \boldsymbol{\nabla} \mu_\tau \cdot d_\tau \boldsymbol{z}  = \partial_\tau \mu_\tau +  \boldsymbol{\nabla} \mu_\tau \cdot (\boldsymbol{x} -\boldsymbol{y}) \\
	\leq  -\frac{1}{2} \|\boldsymbol{\nabla} \mu_\tau\|^2 + \boldsymbol{\nabla} \mu_\tau \cdot (\boldsymbol{x} -\boldsymbol{y})  \leq \frac{1}{2} \|\boldsymbol{x} -\boldsymbol{y}\|^2,
\end{multline}
where the first inequality is the dynamic bound from Eq. \eqref{sup inf argument}, and the second inequality is obtained by completing the square $\|\boldsymbol{\nabla} \mu_\tau + \boldsymbol{y} -\boldsymbol{x}\|^2 \geq 0$.
Integrating this from $\tau=0$ to $\tau=1$, and using the initial condition $\mu_0(\boldsymbol{y}) = \lambda_0(\boldsymbol{y})$, we obtain
\begin{equation}
	\mu_1(\boldsymbol{x}) \leq  \lambda_0(\boldsymbol{y}) + \frac{1}{2} \|\boldsymbol{x}-\boldsymbol{y} \|^2 .
\end{equation}
This inequality holds for every $\boldsymbol{y}$, therefore we can take the infimum on the right hand side and recognize the Hopf-Lax formula Eq. \eqref{Hopf-Lax} to obtain $\mu_1(\boldsymbol{x})\leq \lambda_1(\boldsymbol{x})$.
This proves that the optimal $\lambda_\tau$ for Eq. \eqref{sup inf argument} uniformly satisfies the Hamilton-Jacobi Eq. \eqref{HJ}.

The optimization problem constraint can be rewritten as
\begin{multline}
  \sup_{\lambda_\tau} \left\{ \int d\boldsymbol{x} \, \left[\lambda_\tau p_\tau\right]_0^1 \, \bigg| \, \partial_\tau \lambda_\tau \leq -\frac{1}{2} \|\boldsymbol{\nabla} \lambda_\tau\|^2  \right\} 
  \\= \sup_{\lambda_\tau} \left\{ \int d\boldsymbol{x} \, \left[\lambda_\tau p_\tau\right]_0^1 \, \bigg| \,  \lambda_1(\boldsymbol{x}) - \lambda_0(\boldsymbol{y}) \leq \frac{1}{2} \|\boldsymbol{x} - \boldsymbol{y}\|^2  \right\},
\end{multline}
because maximizing under either the continuous constraint or the static boundary constraint drives the system to the exact same Hopf-Lax formula Eq. \eqref{Hopf-Lax}.

We account for this saturated bound by introducing a Lagrange multiplier field $\pi(\boldsymbol{x}, \boldsymbol{y})$ as
\begin{widetext}\label{Benamou Wasserstein}
	\begin{multline}
		\sup_{\lambda_\tau} \left\{ \int d\boldsymbol{x} \, \left[\lambda_\tau p_\tau\right]_0^1 \, \bigg| \,  \lambda_1(\boldsymbol{x}) - \lambda_0(\boldsymbol{y}) \leq \frac{1}{2} \|\boldsymbol{x} - \boldsymbol{y}\|^2  \right\} \\
		=\sup_{\lambda_\tau} \left\{  \int d\boldsymbol{x} \, \left[\lambda_\tau p_\tau\right]_0^1 + \inf_{\pi>0} \int d\boldsymbol{x} d\boldsymbol{y} \, \pi(\boldsymbol{x}, \boldsymbol{y}) \left[ \lambda_0(\boldsymbol{y}) - \lambda_1(\boldsymbol{x}) + \frac{1}{2} \|\boldsymbol{x} - \boldsymbol{y}\|^2 \right] \right\} \\
		= \inf_{\pi>0} \sup_{\lambda_\tau} \left\{  \int d\boldsymbol{x} \, \left[\lambda_\tau p_\tau\right]_0^1 + \int d\boldsymbol{x} d\boldsymbol{y} \, \pi(\boldsymbol{x}, \boldsymbol{y}) \left[ \lambda_0(\boldsymbol{y}) - \lambda_1(\boldsymbol{x}) + \frac{1}{2} \|\boldsymbol{x} - \boldsymbol{y}\|^2 \right] \right\} \\
		= \inf_{\pi>0} \sup_{\lambda_\tau} \left\{ \int d\boldsymbol{x} \,  \lambda_1(\boldsymbol{x}) \left[ p_1(\boldsymbol{x}) -  \int d\boldsymbol{y} \, \pi(\boldsymbol{x}, \boldsymbol{y}) \right] - \int d\boldsymbol{y} \,\lambda_0(\boldsymbol{y}) \left[ p_0(\boldsymbol{y}) -  \int d\boldsymbol{x} \, \pi(\boldsymbol{x}, \boldsymbol{y}) \right]  \right. \\ \left.  +  \frac{1}{2} \int d\boldsymbol{x} d\boldsymbol{y} \, \pi(\boldsymbol{x}, \boldsymbol{y}) \|\boldsymbol{x} - \boldsymbol{y}\|^2 \right\} \\
		=  \frac{1}{2} \inf_{\pi>0} \left\{ \int d\boldsymbol{x} d\boldsymbol{y} \, \pi(\boldsymbol{x}, \boldsymbol{y}) \|\boldsymbol{x} - \boldsymbol{y}\|^2 \, \bigg| \, \int d\boldsymbol{y} \, \pi(\boldsymbol{x}, \boldsymbol{y}) = p_1(\boldsymbol{x}), \,\, \int d\boldsymbol{x} \, \pi(\boldsymbol{x}, \boldsymbol{y}) = p_0(\boldsymbol{y}) \right\}  \equiv  \frac{1}{2} \mathcal{W}^2(p_1, p_0) ,
\end{multline}
\end{widetext}
where Sion's minimax theorem was applied to the bilinear forms $\lambda_\tau p_\tau$, and the $\pi$ marginals are forced to match the initial $p_0(\boldsymbol{y})$ and final $p_1(\boldsymbol{x})$ density conditions of the fluid dynamics problem as otherwise the supremum on $\lambda_\tau$ would give a positive diverging contribution that gets then discarded by the infimum on $\pi$. In the last step we recognized the definition of the quadratic Wasserstein distance, with $\mathcal{W}(p_1, p_0) = \mathcal{W}(p, q)$ by symmetry.

This concludes the proof, due to Benamou and Brenier \cite{benamou2000computational, villani2009optimal}, of the equivalence between the static map formulation of the Wasserstein distance and the fluid dynamic formulation of optimal transport.

\subsection{Second derivative along the geodesic}\label{sec:HWI}

In the previous section \ref{sec:Review_Optimal_Transport}, the Benamou-Brenier fluid dynamics formulation of the optimal transport problem  \cite{benamou2000computational, villani2009optimal} has been reviewed. In particular, it was shown that the optimal fluid dynamics follows a continuity equation in terms of a dynamical velocity field $\boldsymbol{\nabla}\lambda_\tau$, given by
\begin{equation}\label{appendix optimal transport}
\begin{cases}
  \partial_\tau p_\tau = - \boldsymbol{\nabla} \cdot (p_\tau \boldsymbol{\nabla}\lambda_\tau), \\
  \partial_\tau \lambda_\tau = -\frac{1}{2} \|\boldsymbol{\nabla} \lambda_\tau\|^2 .
\end{cases}
\end{equation}
where the fictitious time $\tau \in [0,1]$ is used to differentiate optimal transport from the physical dynamics in time $t$. The optimal transport considered here transports the arbitrary physical density $p$ at time $t$ to the stationary distribution of the physical dynamics Eq. \eqref{FP}, $p_1 = p^*$. As in the main text, the subscript $\tau$ denotes a spatial field along the transport path at fixed physical time $t$, e.g. $p_\tau \equiv p(\boldsymbol{x}, t, \tau)$, and it is dropped at $\tau=0$, so that $p \equiv p_0$ is the physical density and $\lambda \equiv \lambda_0$ is the initial field. 
This is the natural choice to relate optimal transport with the physical dynamics.
Correspondingly, we consider the relative entropy from the steady state along the optimal transport path, $D_\tau \equiv D[p_\tau\| p^*] = \int d\boldsymbol{x} \, p_\tau \ln(p_\tau/p^*)$, with $D_0 = D$ and $D_1 = 0$. Its first time derivative along the optimal transport dynamics is computed as
\begin{multline}\label{first derivative OT}
d_\tau D_\tau = \int d\boldsymbol{x} \,(\partial_\tau p_\tau) \ln\left(\frac{p_\tau}{p^*}\right) \\ = -\int d\boldsymbol{x} \, \left[ \boldsymbol{\nabla} \cdot (p_\tau \boldsymbol{\nabla} \lambda_\tau) \right] \ln\left(\frac{p_\tau}{p^*}\right) \\
= \int d\boldsymbol{x} \, p_\tau \boldsymbol{\nabla} \lambda_\tau \cdot \boldsymbol{\nabla} \ln\left(\frac{p_\tau}{p^*}\right)\\
=\int d\boldsymbol{x} \left( \boldsymbol{\nabla} p_\tau \cdot \boldsymbol{\nabla} \lambda_\tau + p_\tau \boldsymbol{\nabla} \lambda_\tau \cdot \frac{\boldsymbol{\nabla} \psi}{T} \right)\\
= \int d\boldsymbol{x} \, p_\tau \left( - \nabla^2 \lambda_\tau + \boldsymbol{\nabla} \lambda_\tau \cdot \frac{\boldsymbol{\nabla} \psi}{T} \right),
\end{multline}
where probability conservation $\int d\boldsymbol{x}\, \partial_\tau p_\tau = 0$ was considered, and integration by parts was performed twice assuming $p_\tau (\boldsymbol{\nabla} \lambda_\tau)  \ln\left(p_\tau/p^*\right)$ and $p_\tau \boldsymbol{\nabla} \lambda_\tau$  vanish fast enough at infinity. 
The second time derivative along the geodesic path is then composed of two terms,
\begin{multline}\label{second_derivative_split}
d_\tau^2 D_\tau = \int d\boldsymbol{x} \, (\partial_\tau p_\tau) \left( -\nabla^2 \lambda_\tau + \boldsymbol{\nabla} \lambda_\tau \cdot \frac{\boldsymbol{\nabla} \psi}{T} \right) \\
+ \int d\boldsymbol{x} \, p_\tau \,  \left( -\nabla^2 \partial_\tau\lambda_\tau + \frac{\boldsymbol{\nabla} \psi}{T} \cdot \boldsymbol{\nabla} \partial_\tau\lambda_\tau \right) .
\end{multline}
The first term is evaluated by substituting the continuity equation from Eq. \eqref{appendix optimal transport} and integrating by parts as
\begin{multline}\label{term_one_ot}
\int d\boldsymbol{x} \, (\partial_\tau p_\tau) \left( -\nabla^2 \lambda_\tau + \boldsymbol{\nabla} \lambda_\tau \cdot \frac{\boldsymbol{\nabla} \psi}{T} \right) \\
= -\int d\boldsymbol{x} \, \left[\boldsymbol{\nabla} \cdot (p_\tau \boldsymbol{\nabla} \lambda_\tau) \right] \left( -\nabla^2 \lambda_\tau + \boldsymbol{\nabla} \lambda_\tau \cdot \frac{\boldsymbol{\nabla} \psi}{T} \right)  \\
= \int d\boldsymbol{x} \, p_\tau \boldsymbol{\nabla} \lambda_\tau \cdot \boldsymbol{\nabla} \left( -\nabla^2 \lambda_\tau + \boldsymbol{\nabla} \lambda_\tau \cdot \frac{\boldsymbol{\nabla} \psi}{T} \right)  \\
= \int d\boldsymbol{x} \, p_\tau \boldsymbol{\nabla} \lambda_\tau  \cdot \Big[ - \boldsymbol{\nabla} \nabla^2 \lambda_\tau \\ + \frac{1}{T}  \left( \boldsymbol{\nabla} \boldsymbol{\nabla} \lambda_\tau \cdot \boldsymbol{\nabla} \psi + \boldsymbol{\nabla} \boldsymbol{\nabla} \psi \cdot \boldsymbol{\nabla} \lambda_\tau \right)\Big] .
\end{multline}

The second term of Eq. \eqref{second_derivative_split} is evaluated by utilizing the Hamilton-Jacobi equation $\partial_\tau \lambda_\tau = -\frac{1}{2} \|\boldsymbol{\nabla} \lambda_\tau\|^2$ of Eq. \eqref{appendix optimal transport} as
\begin{multline}\label{term_two_ot}
 \int d\boldsymbol{x} \, p_\tau \,  \left( -\nabla^2 \partial_\tau\lambda_\tau + \frac{\boldsymbol{\nabla} \psi}{T} \cdot \boldsymbol{\nabla} \partial_\tau\lambda_\tau \right) \\
 = \int d\boldsymbol{x} \, p_\tau \Big[ \|\boldsymbol{\nabla} \boldsymbol{\nabla} \lambda_\tau\|^2 + \boldsymbol{\nabla} \lambda_\tau \cdot \boldsymbol{\nabla} \nabla^2 \lambda_\tau \\ - \frac{1}{T} (\boldsymbol{\nabla} \boldsymbol{\nabla} \lambda_\tau \cdot \boldsymbol{\nabla} \lambda_\tau) \cdot \boldsymbol{\nabla} \psi \Big],
\end{multline}
where the vector calculus identities $\boldsymbol{\nabla} \| \boldsymbol{\nabla}\lambda_\tau \|^2= 2 \boldsymbol{\nabla}\boldsymbol{\nabla}\lambda_\tau\cdot \boldsymbol{\nabla}\lambda_\tau$ and $\boldsymbol{\nabla}\cdot (\boldsymbol{\nabla}\boldsymbol{\nabla}\lambda_\tau\cdot \boldsymbol{\nabla}\lambda_\tau) = \|\boldsymbol{\nabla}\boldsymbol{\nabla}\lambda_\tau \|^2 + \boldsymbol{\nabla} \lambda_\tau \cdot \boldsymbol{\nabla} \nabla^2 \lambda_\tau$ were used. By summing the two evaluated components Eqs. \eqref{term_one_ot} and \eqref{term_two_ot}, and considering the Hessian symmetry $(\boldsymbol{\nabla} \boldsymbol{\nabla} \lambda_\tau \cdot \boldsymbol{\nabla} \lambda_\tau) \cdot \boldsymbol{\nabla} \psi =(\boldsymbol{\nabla} \boldsymbol{\nabla} \lambda_\tau \cdot \boldsymbol{\nabla} \psi) \cdot \boldsymbol{\nabla} \lambda_\tau$, the second derivative evaluates to
\begin{equation}\label{second_derivative_ot}
d_\tau^2 D_\tau = \int d\boldsymbol{x} \, p_\tau \left( \|\boldsymbol{\nabla} \boldsymbol{\nabla} \lambda_\tau\|^2 + \boldsymbol{\nabla} \lambda_\tau \cdot \frac{\boldsymbol{\nabla} \boldsymbol{\nabla} \psi}{T} \cdot \boldsymbol{\nabla} \lambda_\tau \right),
\end{equation}
which is main text Eq. (12), and it is just a rewriting of a derivation in Ref. \cite{otto2000generalization} in the formalism adopted here.

\subsection{Entropic refinement}
\label{sec:entropic_refinement}

The nonnegativity bound $\|\boldsymbol{\nabla}\boldsymbol{\nabla}\lambda_\tau\|^2 \geq 0$ used to derive the HWI inequality \cite{otto2000generalization} [Eq. (14) of the main text] can be refined by the trace inequality
\begin{multline}
    \|\boldsymbol{\nabla}\boldsymbol{\nabla}\lambda_\tau\|^2 = \boldsymbol{\nabla}\boldsymbol{\nabla}\lambda_\tau\cdot \boldsymbol{\nabla}\boldsymbol{\nabla}\lambda_\tau 
    = \text{Tr}\left[ (\boldsymbol{\nabla}\boldsymbol{\nabla}\lambda_\tau)^2  \right] \\\geq \frac{1}{n} \left[ \text{Tr}(\boldsymbol{\nabla}\boldsymbol{\nabla}\lambda_\tau) \right]^2 
    = \frac{1}{n} \left( \nabla^2 \lambda_\tau \right)^2 ,
\end{multline}
where $n$ is the number of space dimensions.
By further applying Jensen's inequality to the corresponding integral in Eq. \eqref{second_derivative_ot} yields
\begin{multline}\label{entropic majorization}
    \int d\boldsymbol{x}\, p_\tau \|\boldsymbol{\nabla}\boldsymbol{\nabla}\lambda_\tau\|^2 
    \geq \frac{1}{n}  \int d\boldsymbol{x}\, p_\tau \left( \nabla^2 \lambda_\tau \right)^2 \\
    \geq \frac{1}{n}  \left( \int d\boldsymbol{x}\,  p_\tau \nabla^2 \lambda_\tau \right)^2 = \frac{1}{n} \left( d_\tau S_\tau \right)^2,
\end{multline}
where the last equality yields the dynamics of the entropy $S_\tau \equiv -\int d\boldsymbol{x}\, p_\tau \ln p_\tau$ along the optimal transport path. 
The corresponding refinement of the second derivative lower bound $d^2_\tau D_\tau \geq (\kappa/T) \mathcal{W}^2$, with $\mathcal{W} \equiv \mathcal{W}(p, p^*)$ as in the main text, reads
\begin{equation}\label{refinement d^2_t D}
    d^2_\tau D_\tau \geq \kappa \frac{\mathcal{W}^2}{T}  +\frac{1}{n} \left( d_\tau S_\tau \right)^2,
\end{equation}
which is analogous to the so-called curvature-dimension condition \cite{erbar2015equivalence}.
Integrating this additional term as in Eq. (11) of the main text, one obtains the refined HWI inequality
\begin{multline}\label{HWI}
    \sqrt{-d_t D}  \geq \frac{\mathcal{W}}{\sqrt{T}} \frac{\kappa}{2}  \\ + \frac{\sqrt{T}}{\mathcal{W}}  \left( D + \frac{1}{n} \int_0^1 d\tau \, (1-\tau) \left( d_\tau S_\tau \right)^2 \right)  ,
\end{multline}
which accounts for the entropy variation along the optimal transport path.

\subsection{Gaussian case and saturation}

To analytically investigate the tightness of the Otto--Villani thermodynamic bounds, we consider a 1D overdamped Brownian particle in a harmonic trap and the two quenches between the stiffnesses $\gamma>\kappa$ discussed in the main text: the expansion $\gamma\to\kappa$, in which the system is initially in thermal equilibrium in the steeper trap and relaxes in the softer one, and the reverse quench $\kappa\to\gamma$, in which the density is compressed. We denote by $k$ the stiffness of the final trap, $\psi(x)=\frac{1}{2}kx^2$, so that $k=\kappa$ for the expansion and $k=\gamma$ for the compression. Since $\partial_x^2\psi=k$ is constant, $k$ is also the minimum convexity, Eq.~(13) of the main text, that enters the HWI.

Because the potential is harmonic and the initial state is thermal, the probability density remains Gaussian at all times, $p(x,t)=\mathcal{N}(0,\xi(t))$, and relaxes to $p^*(x)=\mathcal{N}(0,\xi^*)$ with variance $\xi^*=T/k$. From the Fokker-Planck equation, the variance evolves according to $d_t\xi=2(T-k\xi)$. In terms of the dimensionless variance ratio $\omega(t)\equiv\xi(t)/\xi^*$ this reads $d_t\omega=2k(1-\omega)$, with solution $\omega(t)=1+[\omega(0)-1]\,e^{-2kt}$, where $\omega(0)=\kappa/\gamma<1$ for the expansion and $\omega(0)=\gamma/\kappa>1$ for the compression. Hence $\omega(t)<1$ throughout the expansion and $\omega(t)>1$ throughout the compression.

The KL divergence between the two centered Gaussians, the relaxation speed, and the Wasserstein distance evaluate to
\begin{equation}\label{eq:KL_gaussian}
D = \frac{1}{2}(\omega - 1 - \ln \omega),
\end{equation}
\begin{equation}\label{dtD_gaussian}
\sqrt{-d_t D} = \sqrt{k}\,\frac{|1-\omega|}{\sqrt{\omega}},
\end{equation}
\begin{equation}\label{W_gaussian}
\mathcal{W} = \sqrt{\frac{T}{k}}\,|1-\sqrt{\omega}|.
\end{equation}

Along the Benamou-Brenier geodesic from $p$ to $p^*$, the variance of $p_\tau$ is $\xi_\tau = [\sqrt{\xi} + \tau(\sqrt{\xi^*} - \sqrt{\xi})]^2$, with $\xi \equiv \xi_0 = \xi(t)$, and the velocity field is linear in space,
\begin{equation}
    \partial_x \lambda_\tau = \frac{\sqrt{\xi^*} - \sqrt{\xi}}{\sqrt{\xi_\tau}}\, x ,
\end{equation}
so that $\partial_x^2\lambda_\tau$ is uniform in space, with $\partial_x^2 \lambda = 1/\sqrt{\omega} -1$ at $\tau=0$.
The physical local mean velocity $-T\partial_x\ln(p/p^*)=T(1/\xi-1/\xi^*)\,x$ is also linear in $x$. The two velocity fields are therefore globally proportional, and the general bound of Eq.~(10) of the main text,
\begin{equation}\label{general_bound}
\sqrt{-d_t D} \geq \frac{\sqrt{T}}{\mathcal{W}} \left| (d_\tau D_\tau)_{\tau=0}\right| ,
\end{equation}
holds as an equality for both quenches at all times. Explicitly,
\begin{multline}\label{slope_gaussian}
(d_\tau D_\tau)_{\tau=0} = \int dx \, p\left[-\partial_x^2\lambda + (\partial_x\lambda) \frac{\partial_x\psi}{T}\right]\\
= -\frac{(1-\omega)(1-\sqrt{\omega})}{\sqrt{\omega}},
\end{multline}
which together with Eqs.~\eqref{dtD_gaussian} and \eqref{W_gaussian} saturates Eq.~\eqref{general_bound}. Since $1-\omega$ and $1-\sqrt{\omega}$ always have the same sign, $(d_\tau D_\tau)_{\tau=0}<0$ for any $\omega\neq 1$, so that the absolute-value majorization leading to the HWI is also tight.

Since $\int dx\, p_\tau(\partial_x\lambda_\tau)^2=\mathcal{W}^2$ at all transport times and $\partial_x^2\psi=k$, the second derivative of Eq.~(12) of the main text splits as
\begin{equation}\label{d2D_gaussian}
d^2_\tau D_\tau = \frac{k\mathcal{W}^2}{T} + \left(\partial_x^2\lambda_\tau\right)^2 .
\end{equation}
As $\partial_x^2\psi=k$ everywhere, the minimum-convexity majorization is tight, and the energetic term $k\mathcal{W}^2/T=(1-\sqrt{\omega})^2$ is constant along the geodesic. The entropic term $(\partial_x^2\lambda_\tau)^2=(\sqrt{\xi^*}-\sqrt{\xi})^2/\xi_\tau$ is largest at the narrower end of the geodesic, and it equals $(d_\tau S_\tau)^2$, since in 1D with $\partial_x^2\lambda_\tau$ uniform in space the trace and Jensen inequalities of Sec.~\ref{sec:entropic_refinement} hold as equalities. Its weighted and unweighted integrals along the geodesic are
\begin{equation}\label{E_gaussian}
\int_0^1 d\tau\,(1-\tau)\left(\partial_x^2\lambda_\tau\right)^2 = \frac{1}{\sqrt{\omega}}-1+\frac{1}{2}\ln\omega ,
\end{equation}
\begin{equation}\label{G_gaussian}
\int_0^1 d\tau\left(\partial_x^2\lambda_\tau\right)^2 = \frac{(1-\sqrt{\omega})^2}{\sqrt{\omega}} .
\end{equation}
Consistently, the exact slope $-(d_\tau D_\tau)_{\tau=0}=\int_0^1 d\tau\, d^2_\tau D_\tau$, which follows from $(d_\tau D_\tau)_{\tau=1}=0$, is the sum of the energetic term $(1-\sqrt{\omega})^2$ and Eq.~\eqref{G_gaussian}, in agreement with Eq.~\eqref{slope_gaussian}.

The only non-saturated step leading to the HWI [Eq.~(14) of the main text] is therefore the Frobenius norm majorization, which neglects the entropic term. The HWI reads here
\begin{multline}\label{HWI_gaussian}
\sqrt{-d_t D} \geq  \frac{\sqrt{T}}{\mathcal{W}}D + \frac{k}{2}\frac{\mathcal{W}}{\sqrt{T}}\\
= \frac{\sqrt{k}}{2}\,\frac{2\omega - 2\sqrt{\omega} - \ln \omega}{|1-\sqrt{\omega}|},
\end{multline}
and its gap is exactly the neglected entropic term,
\begin{multline}\label{HWI_gap_gaussian}
\sqrt{-d_t D} - \frac{\sqrt{T}}{\mathcal{W}}D - \frac{k}{2}\frac{\mathcal{W}}{\sqrt{T}} \\
= \frac{\sqrt{T}}{\mathcal{W}}\int_0^1 d\tau\,(1-\tau)\left(d_\tau S_\tau\right)^2\\
= \sqrt{k}\,\frac{1/\sqrt{\omega}-1+\frac{1}{2}\ln\omega}{|1-\sqrt{\omega}|} .
\end{multline}
This is strictly positive for $\omega\neq 1$, as the entropy changes along the geodesic, so that the HWI is never saturated here, whereas the refined HWI of Eq.~\eqref{HWI} with $n=1$, which retains exactly this term, is saturated.

Since Eq.~\eqref{general_bound} is saturated, the ratio of the HWI to the exact relaxation speed equals one minus the ratio of Eq.~\eqref{E_gaussian} to the exact slope. With $a\equiv\sqrt{\omega}$ it reads
\begin{equation}\label{R_gaussian}
R(a)=\frac{a\,(a^2-a-\ln a)}{(1-a)^2(1+a)} ,
\end{equation}
which is independent of $k$ and $T$, so that comparing the two quenches reduces to comparing values of $\omega$.

At the quench, the two processes traverse the same geodesic, between $\mathcal{N}(0,T/\gamma)$ and $\mathcal{N}(0,T/\kappa)$, in opposite directions, with $\omega(0)=\kappa/\gamma$ for the expansion and $\omega(0)=\gamma/\kappa$ for the compression. Reversing the direction exchanges the endpoints and thus maps $a\to 1/a$. The unweighted integral of Eq.~\eqref{G_gaussian} is invariant under this map, whereas the energetic term is larger for the compression by the factor $\gamma/\kappa$, so that the exact slope of Eq.~\eqref{slope_gaussian} is larger for the compression, by the factor $\sqrt{\gamma/\kappa}$. Moreover, the weighted integral of Eq.~\eqref{E_gaussian} is larger for the expansion, which starts at the narrower end of the geodesic where the entropic term is largest: the difference between its values at $a<1$ and $1/a$, namely $1/a-a+2\ln a$, vanishes at $a=1$ and has derivative $-(1/a-1)^2\leq 0$. Both effects make the relative gap of the HWI larger for the expansion.

\begin{figure}[ht]
    \centering
    \includegraphics[width=\columnwidth]{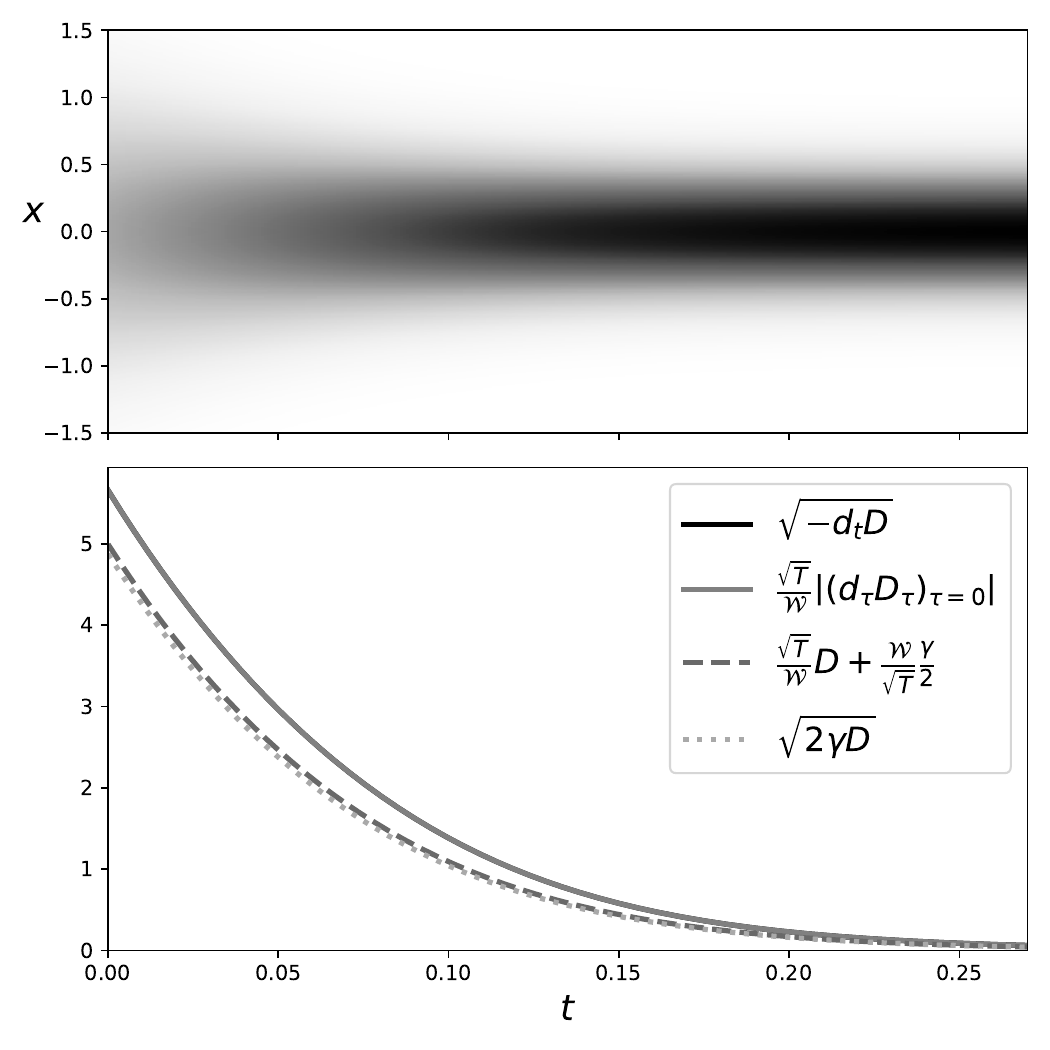}
\caption{Thermal compression in a harmonic potential, the reverse quench $\kappa\to\gamma$ of the expansion in [Main Text Fig.~(4)], with the same parameters $T=1$, $\kappa=2$, and $\gamma=10$. The system is initially in thermal equilibrium in the trap of stiffness $\kappa$ and relaxes in the trap of stiffness $\gamma$, so that the HWI [Eq.~(14) of the main text] is evaluated with the minimum convexity of the final trap, $k=\gamma=10$.}
    \label{Gauss_contraction}
\end{figure}

This ordering holds at all times. The derivative of Eq.~\eqref{R_gaussian} is $R'(a)=M(a)/[(a-1)^3(1+a)^2]$, with
\begin{equation}
M(a)=(1+a+2a^2)\ln a+1+2a-3a^2 .
\end{equation}
Since $M(1)=M'(1)=M''(1)=0$, Taylor's theorem gives $M(a)=M'''(c)\,(a-1)^3/6$ for some $c$ between $1$ and $a$, and since $M'''(c)=(4c^2-c+2)/c^3>0$ it follows that $R'(a)=M'''(c)/[6(1+a)^2]>0$ for all $a$. The ratio therefore increases strictly from $0$ for $\omega\to0$, through $3/4$ at $\omega=1$, to $1$ for $\omega\to\infty$. As $\omega(t)<1$ throughout the expansion and $\omega(t)>1$ throughout the compression, the HWI captures less than $3/4$ of the relaxation speed at all times in the former and more than $3/4$ in the latter, see Fig.~\ref{Gauss_contraction}.

\subsection{Thermodynamic speed limits.}

The time-integrated excess entropy production is written $\sigma_{[0, t_f]}  =  D[p(0)\| p^*] -  D[p(t_f)\| p^*]  $, with respect to a finite time interval $t_f$. 
From the LSI it follows $\sigma_{[0, t_f]} \geq D[p(0)\| p^*] \{1-\exp(-2\kappa t_f)\}$, as already observed in Ref. \cite{bao2025universal}.
The time-integrated lower bound corresponding to the general Otto-Villani instantaneous bound [Eq. (10) in the main text] is written
\begin{multline}\label{integrated general}
    \sigma_{[0, t_f]} \geq T\int_0^{t_f} dt\,\frac{\left| (d_\tau D_\tau)_{\tau=0}\right|^2   }{\mathcal{W}^2} \\
    \geq \frac{T}{t_f \,\max_{t\in [0, t_f]}\mathcal{W}^2}  \left| \int_0^{t_f} dt\, (d_\tau D_\tau)_{\tau=0}\right|^2 ,
\end{multline}
where $D_\tau$ and $\mathcal{W} = \mathcal{W}(p(t), p^*)$ depend on $t$ through the physical density, and the last inequality follows from the majorization $\max_{t\in [0, t_f]}\mathcal{W}^2\geq \mathcal{W}^2$ and Jensen's inequality.

These time-integrated bounds can be compared with the Shiraishi-Saito bound \cite{shiraishi2019information},
\begin{equation}\label{Shiraishi}
    \sigma_{[0, t_f]} \geq D[p(0) \|  p(t_f)] \geq 0,
\end{equation}
which is an information-theoretic bound independent of optimal transport.
The bound Eq. \eqref{integrated general} dominates in the small timescales limit $t_f\rightarrow 0$ as the Shiraishi-Saito bound has a scaling of $\mathcal{O}(t_f^2)$ typical of the Fisher information \cite{amari2016information} compared to the $\mathcal{O}(t_f)$ scaling of Eq. \eqref{integrated general} and also of HWI and LSI.
Note that in the opposite limit $t_f\rightarrow \infty$ the Shiraishi-Saito bound converges to the entropy production definition itself.

For overdamped dynamics \cite{seifert2012stochastic}, the total entropy production of any process with local mean velocity $\boldsymbol{v}$ reads $\Sigma_{[0, t_f]} = T^{-1}\int_0^{t_f} dt \int d\boldsymbol{x}\, p\, \|\boldsymbol{v}\|^2$.
After rescaling time as $\tau = t/t_f$, the Benamou-Brenier formulation of Sec.~\ref{sec:Review_Optimal_Transport} implies $\Sigma_{[0, t_f]} \geq \mathcal{W}^2(p(0), p(t_f))/(t_f T)$ for any driving that takes $p(0)$ to $p(t_f)$ within a time $t_f$.
Equality is reached by the time-dependent potential $U(\boldsymbol{x}, t) = -\lambda_{t/t_f}/t_f - T\ln p_{t/t_f}$, where $p_\tau$ and $\lambda_\tau$ solve the optimal transport problem between $p_0 = p(0)$ and $p_1 = p(t_f)$, since the density then evolves along the geodesic $p_{t/t_f}$ with local mean velocity $-\boldsymbol{\nabla}(U + T\ln p_{t/t_f}) = \boldsymbol{\nabla}\lambda_{t/t_f}/t_f$.
The relaxation at fixed potential $\psi$ is one particular, generally non-optimal, driving between the same densities, and in the absence of nonconservative forces its entropy production equals the excess one, $\Sigma_{[0, t_f]} = \sigma_{[0, t_f]}$, while in general $\Sigma_{[0, t_f]} \geq \sigma_{[0, t_f]}$.
For conservative relaxation, this yields the bound of Aurell et al.~\cite{aurell2012refined},
\begin{equation}\label{aurell}
    \sigma_{[0, t_f]} \geq \frac{\mathcal{W}^2(p(0), p(t_f))}{t_f \, T} .
\end{equation}
At finite $t_f$, this is a genuine lower bound because the physical path
generally differs from the constant-speed geodesic between $p(0)$ and
$p(t_f)$. 
In the absence of nonconservative forces, this difference vanishes in the short-time limit, since the local mean velocity $-T\boldsymbol{\nabla}\ln(p/p^*)$ is then a gradient field, which is the minimal-energy velocity field generating the same change of density and thus coincides with the velocity of the corresponding infinitesimal optimal transport~\cite{nakazato2021geometrical,ito2024geometric}.
Applied to an interval
$[t,t+\delta t]$, Eq.~\eqref{aurell} is therefore saturated for $\delta t\to 0$, where
it becomes
\begin{equation}
-d_tD=\frac{1}{T}\left[\lim_{\delta t\to 0}
\frac{\mathcal{W}\big(p(t),p(t+\delta t)\big)}{\delta t}\right]^2 .
\label{eq:Wspeed}
\end{equation}
This is not a lower bound on the relaxation speed, but the definition of the
entropy production rate $T^{-1}\int d\bm{x}\,p\|\bm{v}\|^2$ expressed through
the Wasserstein speed of the physical density.

In the long-time limit $t_f\to\infty$,
instead, the bound of Eq.~\eqref{aurell} vanishes given the convergence
$\mathcal{W}(p(0),p(t_f))\to\mathcal{W}(p(0),p^*)$. For intermediate times it
gives a meaningful geometric lower bound which quantifies the cost of
relaxation at fixed potential compared to the time-dependent optimal protocol.

A nontrivial bound on the \textit{instantaneous} relaxation speed requires instead a
comparison between the direction of the physical flow and a reference
direction. In the Otto--Villani framework, this is the direction of the
geodesic from $p(t)$ toward the steady state $p^*$, with initial velocity
field $\bm{\nabla}\lambda$ as in Sec.~\ref{sec:HWI}.

\bibliography{bibliography}